\documentclass[aps,prb,reprint,superscriptaddress]{revtex4-1}

\usepackage[colorlinks=true,citecolor=blue,linkcolor=blue]{hyperref}
\usepackage{graphicx}

\begin{document}

\title{Impact of intra-grain spin wave reflections on nano-contact spin torque oscillators}

\author{Anders J. Eklund}
\email{a.j.eklund@fys.uio.no}
\affiliation{Department of Physics, University of Oslo, Box 1048 Blindern, 0316 Oslo, Norway}

\author{Mykola Dvornik}
\affiliation{Department of Physics, University of Gothenburg, 412 96 Gothenburg, Sweden}

\author{Fatjon Qejvanaj}
\affiliation{NanOsc AB, Electrum 205, 164 40 Kista, Sweden}

\author{Sheng Jiang}
\affiliation{Department of Applied Physics, School of Engineering Sciences, KTH Royal Institute of Technology, Electrum 229, 164 40 Kista, Sweden}

\author{Sunjae Chung}
\affiliation{Department of Physics, University of Gothenburg, 412 96 Gothenburg, Sweden}
\affiliation{Department of Physics Education, Korea National University of Education, Cheongju 28173, Korea}

\author{Johan {\AA}kerman}
\affiliation{Department of Physics, University of Gothenburg, 412 96 Gothenburg, Sweden}
\affiliation{NanOsc AB, Electrum 205, 164 40 Kista, Sweden}
\affiliation{Department of Applied Physics, School of Engineering Sciences, KTH Royal Institute of Technology, Electrum 229, 164 40 Kista, Sweden}

\author{B. Gunnar Malm}
\affiliation{Division of Electronics and Embedded Systems, School of Electrical Engineering and Computer Science, KTH Royal Institute of Technology, Electrum 229, 164 40 Kista, Sweden}

\date{August 19, 2020}

\begin{abstract}
We investigate the origin of the experimentally observed varying current-frequency nonlinearity of the propagating spin wave mode in nano-contact spin torque oscillators. Nominally identical devices with 100 nm diameter are characterized by electrical microwave measurements and show large variation in the generated frequency as a function of drive current. This quantitative and qualitative device-to-device variation is described in terms of continuous and discontinuous nonlinear transitions between linear current intervals.  The thin film grain microstructure in our samples is determined using atomic force and scanning electron microscopy to be on the scale of 30 nm. Micromagnetic simulations show that the reflection of spin waves against the grain boundaries results in standing wave resonance configurations. For a simulated device with a single artificial grain, the frequency increases linearly with the drive current until the decreased wavelength eventually forces another spin wave anti-node to be formed. This transition results in a discontinuous step in the frequency versus current relation. Simulations of complete, randomly generated grain microstructures additionally shows continuous nonlinearity and a resulting device-to-device variation in frequency that is similar to the experimental levels. The impact of temperature from 4 K to 300 K on the resonance mode-transition nonlinearity and frequency noise is investigated using simulations and it is found that the peak levels of the spectral linewidth as a function of drive current agrees quantitatively with typical levels found in experiments at room temperature.

\end{abstract}


\maketitle

\section{Introduction}
The nano-contact spin torque oscillator \cite{Silva2008,Chen2016} (NC-STO) is a spintronic microwave oscillator in which the spin transfer torque \cite{Slonczewski1996a,Berger1996,Ralph2008a} (STT), induced by an electrical direct current, counteracts the Gilbert damping and enables a persistent precession of the magnetization in the free layer. Through the giant magnetoresistance effect \cite{Baibich1988,Binasch1989}, this precessing magnetization direction results in a correspondingly time-varying device resistance, which together with the dc drive current produces an oscillating voltage signal. The excitation of the magnetic free layer can take place in the form of a circularly trajecting magnetic vortex \cite{Pufall2007,Mistral2008} at lower frequencies (from hundreds of MHz to a couple of GHz) and, on the order of tens of GHz \cite{Bonetti2009}, as various spin wave modes. These spin wave modes include the solitonic "bullet" mode \cite{Slavin2005} and propagating spin wave mode \cite{Slonczewski1999a,Madami2011,Madami2015} for systems with an in-plane magnetic anisotropy for the free layer.

Out of these high-frequency modes, the propagating spin wave mode possesses several features that makes it more attractive for applications. First, it can be excited exclusively while the bullet mode can only be excited in conditions where also the propagating mode exists \cite{Bonetti2010}. Simultaneous excitability with the possibility of mode-hopping between these modes accounts for the comparatively high frequency-domain linewidths observed during these conditions. Second, its propagating nature increases the size of the oscillating system and thereby increases the frequency stability. Third, it blue-shifts with the application of increased drive current magnitude and thereby provides higher frequencies. Apart from the high frequency range and comparatively high frequency stability most useful in microwave RF applications, the large-amplitude propagating spin waves are also attractive for use in magnonic circuits \cite{Kruglyak2010,Bonetti2013,Chumak2015}.

One general property of the propagating spin wave mode that has so far not been explained or modelled, is the rich variety of features in the frequency versus current behavior $f(I_{\text{DC}})$, which generally shows regions of linear dependence joined by nonlinear transitions\cite{Rippard2006,Tamaru2012}. These nonlinear transitions can be either continuous or discontinuous, i.e., appear as $f(I_{\text{DC}})$ bending or discrete steps in frequency. It may be considered natural to describe the linear regions as sub-modes of the propagating spin wave mode, but a physical mechanism behind such a degeneracy has not yet been reported. The frequency steps are of the order of 1 GHz and cannot be explained as the much larger $\sim$10 GHz steps between the different higher-order Slonczewski modes \cite{Houshang2018}. We would like to point out that the phenomenon under investigation is a higher order nonlinearity not to be confused with the general auto-oscillator nonlinearity property \cite{Slavin2009a}, which concerns the presence of coupling between the amplitude and frequency of the magnetization precession. While the amplitude-frequency coupling is the mechanism that makes it possible to tune the STO frequency by changing the magnitude of the drive current $I_{\text{DC}}$, we see no physical reason why this coupling by itself should give rise to the type of complexity found in the experimentally measured $f(I_{\text{DC}})$ behavior.

The $f(I_{\text{DC}})$ nonlinearity is of direct interest for any tehnological application of the NC-STO for two main reasons: frequency stability (phase noise) and device-to-device variability. Within the nonlinear transition intervals, the frequency stability is decreased which is commonly observed as an increase in the spectral linewidth \cite{Rippard2006,Eklund2014}. More detailed measurements have shown that the $f(I_{\text{DC}})$ nonlinearity is associated with increased levels of both white and $1/f$ frequency noise \cite{Eklund2014}. Nominally identical devices also differ significantly in the position and type of the $f(I_{\text{DC}})$ nonlinearity \cite{Tamaru2012}, which translates to device-to-device variation in $f(I_{\text{DC}})$. The same type of qualitative and quantitative frequency variability that is found between devices can also be seen when changing the angle of the in-plane component\cite{Tamaru2012,Pufall2012} or polarity\cite{Pufall2012} of the applied magnetic field. These studies have concluded that the measured behavior is consistent with an oscillation that takes place in magnetic "hotspots"\cite{Tamaru2012} or "subregions"\cite{Pufall2012} defined by an inhomogeneous effective magnetic field and/or spin polarization ratio originating from microstructural inhomogeneity. The linearity and nonlinearity of $f(I_{\text{DC}})$ would in this context arise due to a complex interplay between these different subregions. Although this is certainly a possible scenario, we consider it less likely that the selection of the dominating subregion would be so sensitive to the drive current $I_{\text{DC}}$. We do not see a clear reason why this selection would change and would rather expect that the frequency would be set by the same subregion throughout most of the operating current range.

When setting up our simulations in an attempt to recreate the nonlinearity, we initially noted the sensitivity to the boundary conditions of the simulation space. In particular, we found that periodic boundary conditions in combination with a simulation space being about one order of magnitude larger than the nano-contact resulted in discontinuous steps in $f(I_{\text{DC}})$. In this configuration, the spin wave propagated from the NC to any of the simulation space borders, re-entered from the opposing border and still had a notable amplitude as it re-entered the NC area. This led us into the conceptually simpler hypothesis that the $f(I_{\text{DC}})$ nonlinearity originates from spin wave propagation and STO self-interaction. Microstructural inhomogeneity and spin wave reflection in combination with axial asymmetry in the oscillation mode would be consistent with both the device-to-device variability and in-plane magnetic field dependence. Micromagnetic simulations have shown that for applied magnetic fields having an in-plane component, the inclusion of the current-induced Oersted magnetic field indeed breaks the symmetry of the propagating mode and propagation instead takes the form of a directed spin wave beam \cite{Hoefer2008,Dumas2013}.

In this work, we investigate the microstructure of the thin film in terms of the size of the metal grains and perform micromagnetic simulations with included grain boundaries. With reduced magnetic exchange coupling at the grain boundaries, the propagating spin wave becomes reflected and travels back to the active region. By self-locking, the spin wave reflections result in resonating spin wave paths that each depends on the distance to the reflecting grain boundary and the wavelength. This leads to multiple sets of resonance frequencies for the different reflecting grain boundaries and provides a direct physical model for the $f(I_{\text{DC}})$ sub-modes and their associated variability. As will be shown, the model is able to recreate both continuous and discontinuous $f(I_{\text{DC}})$ nonlinearity, the device-to-device variability (with reasonable quantitative agreement) and the correct variation of the spectral linewidth.

\section{Experimental Methods}
The samples were fabricated by sputter deposition to form the stack Si/SiO$_{\text{x}}$/\allowbreak Pd8/Cu30/\allowbreak Co8/Cu8/NiFe4.5/\allowbreak Cu3/Pd3 (thicknesses in nm). The film was then patterned into $16 \times 8$ $\mu$m$^2$ mesas by optical lithography and lift-off, followed by sputter deposition of a 30 nm SiO$_2$ insulating layer. Circular nano-contacts with diameter $d_{\text{NC}} = \textrm{100 nm}$ were patterned using electron beam lithography and etched using reactive ion etching. The nano-contact vias were then metallized by Cu during the deposition of the Cu1000/Au400 top contact, which was also defined by optical lithography and lift-off. The top contact has a coplanar waveguide ground-signal-ground (GSG) configuration, where the S pad is connected to the nano-contact and the G pads are connected to the outer regions of the mesa through two $2 \times 4$ $\mu$m$^2$ vias.

The electrical microwave measurements were performed using a 40 GHz-rated GSG microwave probe followed by a bias-T and a 20--40 GHz low-noise amplifier (gain 28 dB, noise figure of 3.0 dB) before recorded on a spectrum analyzer. The bias current was supplied by a Keithley 6221 precision current source, with the positive current direction defined as electrons flowing from the free NiFe layer to the fixed Co layer. The sample and microwave circuit were mounted on an electrically controlled rotating holder with the sample positioned inside the pole gap of an electromagnet. The current driving the electromagnet was feedback-controlled using a PI controller with a calibrated Hall sensor positioned at the center of one of the poles. In all measurements presented in this work, the strength and angle of the applied field (away from the film plane) are $H_{\text{ext}} = \textrm{10.0 kOe}$ and $\theta_{\text{ext}} = 70^\circ$. This field was selected based on Ref. \citenum{Bonetti2012} to optimize the trade-off balance between oscillation power and frequency stability (spectral linewidth), well above the critical angle of $\theta_{\text{ext},c} = 58^\circ$ under which also the localized bullet mode is co-existingly excited.

Atomic force and scanning electron microscopy (AFM, SEM) were performed on a separately prepared, unpatterned Si/SiO$_{\text{x}}$/\allowbreak Pd8/Cu30/\allowbreak Co8/Cu8/NiFe4.5/Ta3 film to capture the structure of the NiFe free layer. AFM was conducted with a JPK NanoWizard 3 NanoScience microscope in the AC tapping mode using an AppNano ACTA tip with a (nominal) radius of curvature of 6 nm. The SEM measurement was performed using the in-lens detector of a Zeiss Ultra 55 microscope at $\sim$3 mm working distance, with a specified resolution of 1.6 nm at 1 kV accelerating voltage.

Ferromagnetic resonance (FMR) measurements of the same unpatterned films, using a NanOsc Instruments PhaseFMR-40, gave the saturation magnetization and Gilbert damping for the NiFe free layer of $4\pi M_{s,\text{NiFe}} = 10.1$ kG, $\alpha_{G,\text{NiFe}} = 0.0135$ and Co fixed layer $4\pi M_{s,\text{Co}} = 19.8$ kG, $\alpha_{G,\text{Co}} = 0.0088$.

\section{Simulation Methods}
Simulations were performed using the open-source, GPU-accelerated micromagnetic simulation package \cite{VansteenKiste2014} mumax$^3$ . For the homogeneous free layer simulations, a $512 \times 512 \times 1$ quadratic grid was used with a cell size of $2.5 \times 2.5 \times 4$ nm$^3$ for the $1280 \times 1280 \times 4$ nm$^3$ free layer representation. The fixed layer and spacer layer were modelled as $1280 \times 1280 \times 8$ nm$^3$. An initial settling step was used to let the full stack relax into its static configuration, taking into account the externally applied field, the dipolar field, the exchange field and the Oersted field. After settling, the fixed layer cells were kept static in order to reduce the computation time. By this approach, the dipolar field of the (static) fixed layer is automatically included in the simulation. Absorbing boundary conditions were implemented similar to those in Ref. \citenum{Bonetti2015} with three encapsulated frames, each with a width of 5 \% of the simulation space, with the damping parameter successively increasing to $\alpha_G = $ 0.05, 0.15 and 0.45. Using these settings and sweeping the current resulted in a frequency versus current behavior free from continuous or discontinuous nonlinearities for the oscillation regime above the threshold, while a linear 50 \% reduction in the simulation space resulted in slight frequency stepping due to wave reflection against the simulation space borders.

The majority of the material parameters for the fully processed samples were taken from our previous work in Ref. \citenum{Dumas2013} with slight adaptation to fit the threshold current and frequency of our experimental sample batch. Selected values were: saturation magnetization $4\pi M_{s,\text{NiFe}} = 8.5$ kG, $4\pi M_{s,\text{Co}} = 17.0$ kG; exchange stiffness $A_{\text{ex,NiFe}} = 1.1\cdot10^{-11}$ J/m, $A_{\text{ex,Co}} = 2.1\cdot10^{-11}$ J/m. For the spin torque, the polarization was 0.3 and the Slonczewski parameter $\Lambda = 1.0$. The Gilbert damping parameter was taken from our FMR measurements:  $\alpha_{G,\text{NiFe}} = 0.0135$ and $\alpha_{G,\text{Co}} = 0.0088$. 

The Oersted field was calculated as that from an infinitely long conductor running down the nano-contact. The magnitude of the field increases linearly from the center of the nano-contact out to the edge, outside which it decays with the inverse distance from the center.

For the simulations including the grain structure, the grains were randomly generated using the Voronoi tesselation extension to mumax$^3$. Using another extension, the exchange coupling across the grain boundaries was reduced by scaling. Tests using a constant scaling factor \cite{LeliaertJAP2014a} between all grains showed well-defined oscillation for 30--100 \% coupling, while 20 \% showed oscillation only for currents below 30 mA and 0--10 \% resulted in broadband noise. Since the exchange coupling is highly sensitive to the inter-atomic distance (it has been calculated \cite{Victora2003} to drop to ~0 already at a distance of 1.5 times the crystalline distance), we consider it more realistic to have a random inter-grain exchange distribution. Knowing little about the grain-to-grain interface structure, we here make a first approximation with a uniform distribution of 0--100 \% coupling. Both the grain tessellation and exchange scaling are set using the same specified seed number for the random number generator, ensuring reproducibility. When simulating the grain structure, no significant difference was observed between the full $1280 \times 1280$ nm$^2$ and halved $640 \times 640$ nm$^2$ simulation spaces, indicating that only an insignificant amount of energy is propagated to the simulation border and back to the active region in this case. This is consistent with spin wave reflection occurring at the grain boundaries having a higher influence than the simulation space border effects. Because of this, the simulations of the grainy free layer were performed using the smaller simulation space in order to reduce the simulation time.

Majority of the simulations were performed at temperature $T = \textrm{300 K}$ using an adaptive \cite{Leliaert2017} timestep which usually settled at around 50 fs. The duration of the simulations were 1 $\mu$s except for the homogeneous film presented in Figure \ref{Fig_Sim_Homogeneous}, where we used 100 ns. For the calculation of the spectral linewidth plotted in Figure \ref{Fig_Sim_Linewidth} the 1 $\mu$s timetrace was split into two and the spectra averaged, resulting in an approximate spectral resolution of 2 MHz.

\section{Results and Discussion}

\subsection{Experiment}

\begin{figure*}[t]
  \begin{center}
    \includegraphics[trim = 0mm 0mm 0mm 0mm, clip]{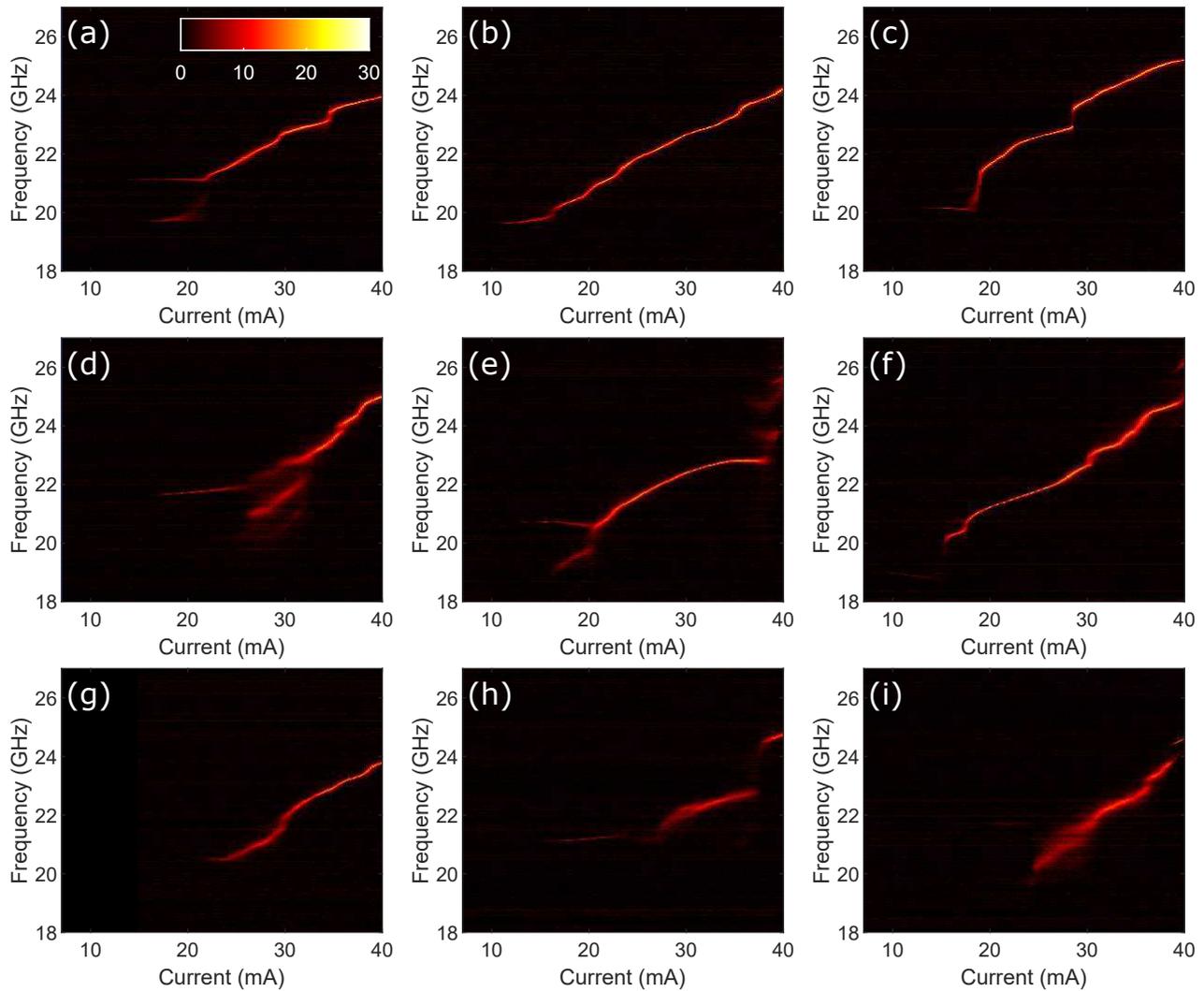}
    \caption[]{(a)--(i) Experimental power spectral density in dB over noise as a function of drive current for nine devices on the same chip, adjacent to each other.}
    \label{Fig_NinePSDs}
  \end{center}
\end{figure*}

We begin by exemplifying the diversity in the device-specific frequency versus current behavior. We do this by measuring nine devices located in a line next to each other, on the same chip. Any wafer-scale manufacturing variability is hence kept at its minimum possible influence. The spectral density as a function of the drive current is shown in Figure \ref{Fig_NinePSDs} for the nine devices. The frequency as a function of current shows a blue-shifting trend for all the devices, consistent with the propagating spin wave mode. At low currents, before the onset of the blue-shifting propagating mode, all the devices show a weaker pre-threshold mode with low, zero or even negative tunability. The pre-threshold mode does in some cases, but not all, connect to the propagating mode. Apart from those general features, it can be said that the behavior differs qualitatively between the devices in terms of the number of simultaneously excited frequencies, the position and height of the discontinuous frequency steps and the linearity or curvature. We refer to both the discontinuous frequency steps and continuously nonlinear current-dependence as nonlinearities, since the simulated behavior for an ideal, homogeneous thin film was found to be highly linear above the threshold current. The devices in Figures \ref{Fig_NinePSDs}(d,h,i) show a particular instability in the low-current section of the propagating mode, with worse defined frequencies.

The devices in Figure \ref{Fig_NinePSDs} show resemblance to previously characterized NC-STOs in similar field configurations, in particular in terms of the presence of linear regions connected by nonlinearities that are either discontinuous or continuous. The diversity among our devices is large but we would like to point out that we here present completely \emph{non-selected} data in its unprocessed form, without reducing it by extracting and displaying only the dominant peak frequencies. The large sample-to-sample variation in the frequency versus current behavior implies that the magnetization dynamics is highly different between the devices. The differences can not easily be explained by device-to-device variation in the nominal parameters such as the film thicknesses since this can not be expected to result in the qualitatively different device characteristics. Nor does such reasoning explain the emergence of the discontinuous and continuous nonlinearities. We identify these nonlinearities and \emph{their} variation as a root cause of the sample-to-sample variation. The (simultaneous) existence of several propagating modes with different frequencies is another open question.

\begin{figure*}[t]
  \begin{center}
	\includegraphics[trim = 0mm 0mm 0mm 0mm, clip]{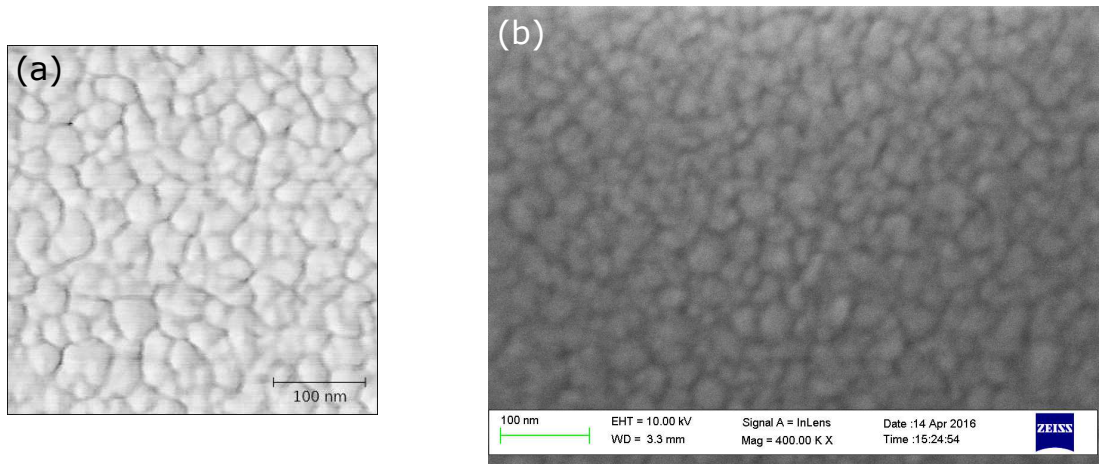}
    \caption[]{(a) Atomic force microscopy (phase contrast mode) and (b) scanning electron microscopy of the surface of the free layer (Si/SiO$_{\text{x}}$/Pd8/Cu30/\allowbreak Co8/Cu8/NiFe4.5/Ta3 stack).}
  	\label{Fig_AFM_SEM}
  \end{center}
\end{figure*}

The large quantitative and qualitative differences between the devices indicate that the origin of the differences is a highly device-specific phenomenon. Figure \ref{Fig_AFM_SEM}(a) shows the AFM measurement of the structure of the free layer film and Figure \ref{Fig_AFM_SEM}(b) shows the SEM measurement of the same film. Both the microscopes reveal a grain structure with grains on the order of 30 nm size. Due to the 4.5 nm small thickness of the free layer, it is reasonable to assume that all the grains are columnar. 

The grain structure constitutes a possible complex source for the similarly complex device-to-device variation and brings up the question of how reasonable the homogeneous-film approximation is for modelling exchange-dominated propagating spin waves. It can be assumed that at the grain boundaries, the exchange interaction may be significantly reduced \cite{Victora2003} with the magnitude of the reduction depending on the individual grain-to-grain interfaces. The random geometry of the grain structure together with the likely random nature of the inter-grain exchange coupling reduction constitute a vast variability space. We next turn to micromagnetic simulations in order to gain insights of the potential effects that the grain structure has on the propagation as well as generation of the spin waves.

\subsection{Simulation: homogeneous thin film}
As a basis for the investigation of the impact of spin wave barriers in the free layer thin film, we first perform simulations of the nominal case of a perfect, homogeneous film.

\begin{figure*}[t]
  \begin{center}
    \includegraphics[trim = 0mm 0mm 0mm 0mm, clip]{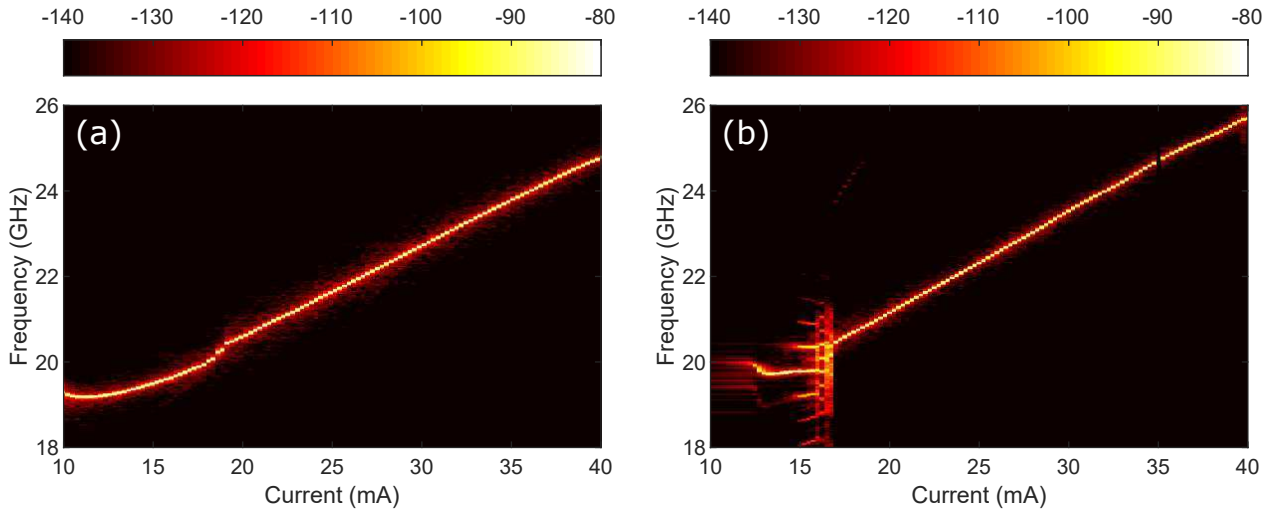}
    \caption[]{Fourier amplitude of $m_y$ (in dB) from simulations with a homogeneous free layer film at $T = \textrm{4 K}$, (a) without and (b) with the Oersted field.}
    \label{Fig_Sim_Homogeneous}
  \end{center}
\end{figure*}

Figure \ref{Fig_Sim_Homogeneous}(a,b) shows the simulated frequency versus current behavior for the nominal case of a perfectly homogeneous free layer film in the cases of exclusion (a) and inclusion (b) of the Oersted field. We find that in the homogeneous film, there is a perfectly linear $f(I_{\text{DC}})$ relation once above 20.4 GHz, i.e. above the onset threshold region of the blue-shifting propagating mode. Above this threshold, the inclusion of the Oersted field does not result in any qualitative difference from the linear $f(I_{\text{DC}})$ behavior. The effect of the Oersted field in the homogeneous-film case is to shift the frequency up by $\sim$0.6 GHz close to the threshold, increasing to $\sim$0.9 GHz at 40 mA.

The linear $f(I_{\text{DC}})$ dependence is an important result, since it shows that there is no inherent mechanism for the magnetization precession or the propagating spin wave mode that introduces nonlinear $f(I_{\text{DC}})$ behavior. In other words, the amplitude-frequency coupling has a constant nonlinearity coefficient within the operating frequency range. This shows that a more complex model of the system is required.

Before the onset of the propagating mode, the so called pre-threshold behavior shows a more dramatic difference. When the Oersted field is included, there is a frequency jump of 0.6 GHz from the pre-threshold mode to the propagating mode whereas there is a continuous transition without the Oersted field. This can be understood by considering the local FMR landscapes. Without the Oersted field, the local FMR frequency is homogeneous and there is no asymmetry that allows different oscillation modes in different regions. This symmetry remains when the drive current is increased and the increased spin torque eventually becomes strong enough to launch the propagating wave. In the case of the Oersted field, there is an asymmetry between the center and the edges of the NC that facilitates the generation of several modes in different volumes. This allows the propagating mode to be excited independently of the pre-threshold mode at a different, higher frequency. As the drive current is increased and the mode volumes grow and start to overlap, the propagating mode eventually extinguishes the pre-threshold mode. The intermodulation products visible in the threshold current range 15--17 mA in Figure \ref{Fig_Sim_Homogeneous}(b) are a result of this coexistence\cite{Dumas2013,Iacocca2015} in time and space of the two modes.

Our result of a high degree of linearity in the frequency as a function of $I_{\text{DC}}$ for the propagating mode is opposite to the result in Ref. \citenum{Puliafito2014}, where nonlinearity was found despite the effort of having implemented spin wave absorbing boundary conditions. However, when we employed identical boundary conditions (in a circle outside the nano-contact) and performed a spatial analysis we did actually observe a certain amount of reflection from the "absorbing" boundary that resulted in an artificially introduced standing spin wave pattern. We we will see in the following sections that reflection and standing spin waves in the physical system is a highly dominant mechanism affecting the frequency selection in a way that introduces the $f(I_{\text{DC}})$ nonlinearity.

\subsection{Simulation: impact of a single barrier}

As a first step towards investigating the possible effects of the grain structure, we simulate the STO behavior in the case where a single barrier is placed in the spin wave path. Being a wave phenomenon, we expect that part of the incident wave is reflected back towards the source, i.e. the active, current-driven region directly below the NC. We set up a barrier in the free layer in form of an artifical rectangular grain with the exchange coupling between it and the surrounding film set to zero. The rectangular grain has a width of 100 nm facing the NC and is 50 nm deep. The NC is at the origin of the $xy$-plane of the sample film; the externally applied field is aligned in the first quadrant of the $xz$-plane and the barrier is positioned along the positive $y$-direction. This is the direction into which the spin wave beam propagates in the case when the Oersted field is included\cite{Dumas2013}; this is the side where the in-plane component of the external field and the Oersted field oppose each other, resulting in a decrease of the local FMR frequency. On the opposite side, the in-plane components add up and bring the local FMR frequency up to a level above the spin wave generation, hence blocking spin wave propagation in that direction.

\begin{figure}[t]
  \begin{center}
	\includegraphics[trim = 0mm 0mm 0mm 0mm, clip]{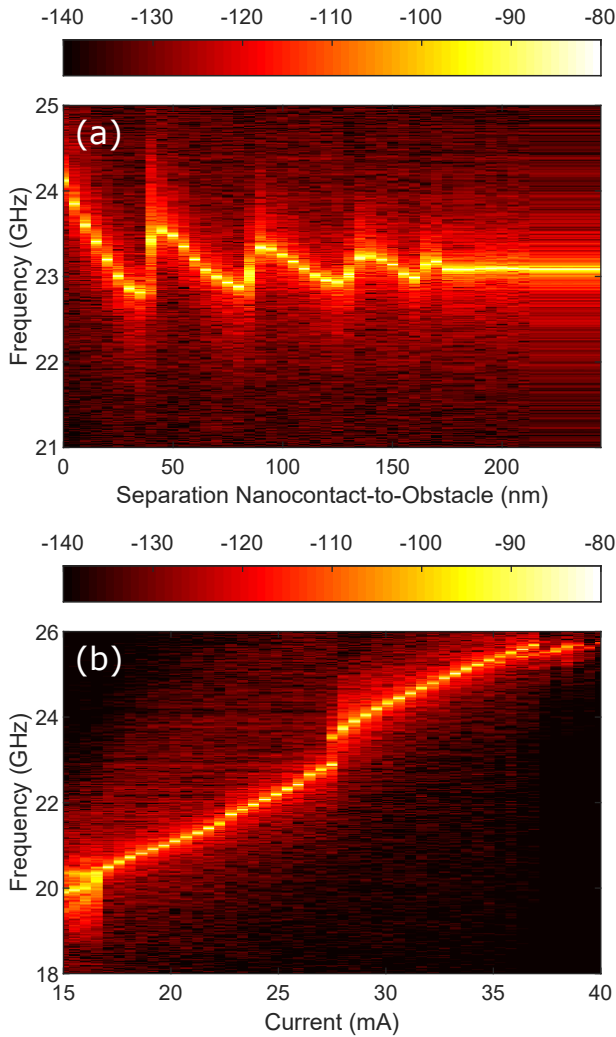}
    \caption[]{Simulations of a barrier with no exchange coupling to the remaining free layer film. Fourier amplitude of $m_y$ (in dB) for: (a) Varied NC-edge to obstacle separation distance $d_{\text{sep}}$ for $I_{\text{DC}} = \textrm{27.75 mA}$. (b) Current sweep for $d_{\text{sep}} = \textrm{125 nm}$.}
    \label{Fig_Sim_Barrier}
  \end{center}
\end{figure}

The result from the simulations with the single, artificial  "wall" grain is shown in Figure \ref{Fig_Sim_Barrier}. Figure \ref{Fig_Sim_Barrier}(a) displays the power spectral density as a function of the edge-to-edge separation distance $d_{\text{sep}}$ between the NC and the wall, for a fixed current of $I_{\text{DC}}$ = 27.75 mA. At this level of drive current, the nominal, ideal film oscillation is stable at 23.1 GHz (see Figure \ref{Fig_Sim_Homogeneous}(b)). Figure \ref{Fig_Sim_Barrier}(a) shows a periodic pattern with a period of 45 nm. This corresponds very well to half of the spin wave length of $85-90 \textrm{ nm}$ which we read off from the simulation space in the unperturbed situation. At its peaks, the wall shifts the oscillation frequency away from the homogeneous-film case frequency up to 1.0 GHz positive and 0.3 GHz negative, where the magnitude of the frequency shift decreases with $d_{\text{sep}}$. The effect of the wall is visible up to $d_{\text{sep}}$ = 210 nm, where the STO once again attains its ideal, homogeneous-film frequency. The downward frequency slope can be understood as the consequence of a forced and gradually enlarged spin wave length as the STO strives for spin wave resonance at a gradually longer distance. The upward jumps in frequency occur when resonance eventually occurs for one additional node and anti-node, which rapidly forces a shorter wavelength. We note that the effect is asymmetric towards higher frequency, corresponding to a preferred situation of shorter wavelength (more nodes). This is a consequence of the general coupling in STOs between the oscillation amplitude and frequency: the frequency increases with the amplitude for the propagating spin wave mode in this magnetic field. The strongest and dominating resonance occurs for the standing wave configuration that has the highest amplitude and, as a consequence, also has the highest frequency, shortest wavelength and highest number of nodes. Larger amplitude and higher frequency is obtained for resonance at small $d_{\text{sep}}$ due to the lower amount of spin wave damping along the shorter propagation path.

Figure \ref{Fig_Sim_Barrier}(b) shows a current sweep for the case of fixed $d_{\text{sep}}$ = 125 nm. The artificially introduced spin wave barrier introduces nonlinearity in $f(I_{\text{DC}})$ similar to the type that is characteristic for the experimental devices; we notice the appearance of a discontinuous frequency step of 0.65 GHz at 27 mA and a small degree of continuous nonlinear behavior both below and above the step. Compared to the homogeneous film case in Figure \ref{Fig_Sim_Homogeneous}(b), the introduction of the barrier generally pushes the frequency higher. This is consistent with the general preference of selecting a higher frequency (i.e. squeezing in an additional standing wave node), as found in Figure \ref{Fig_Sim_Barrier}(a).

This examplifies how a reflected spin wave can alter the generation frequency, and can be considered as STO self-interaction. This is made possible by the ability of the STO to be pulled towards the frequency of an injected signal and phase-lock to it, which has previously been shown to occur both for injected electrical RF signals \cite{Rippard2005,Florez2008,Urazhdin2010,Quinsat2011,muduli2011if,Finocchio2012} and incoming spin waves\cite{Kaka2005,Mancoff2005,Pufall2006,Rezende2007,Sani2013a,Houshang2015}. We find that in the case of spin wave reflection and self-interaction, the wavelength becomes tuned to form a standing wave between the NC and the barrier in a positive feedback loop. The STO will stabilize its oscillation by tuning its frequency such that an anti-phase reflected wave is avoided.

The degree to which the STO adapts the frequency to the available standing wave-frequencies depends on the amplitude of the incoming reflected spin wave in relation to the STO amplitude. If an incoming anti-phase wave is weak enough, the STO frequency will be unaffected. In such a case, the reflected wave merely constitutes a perturbation that can be expected to introduce phase noise but not shift the frequency. In Figure \ref{Fig_Sim_Barrier}(a) we find that the distance where frequency-shifting becomes negligible for our simulated devices is around $d_{\text{sep}} = \textrm{210 nm}$ away from the NC edge.

Similar standing spin wave effects were obtained when the same barrier was left with full exchange coupling across its boundaries and instead the NiFe Gilbert damping parameter $\alpha_G$ was increased to 1.0.

\subsection{Simulation: impact of grain structure}

\begin{figure*}[t]
  \begin{center}  
    \includegraphics[trim = 0mm 0mm 0mm 0mm, clip]{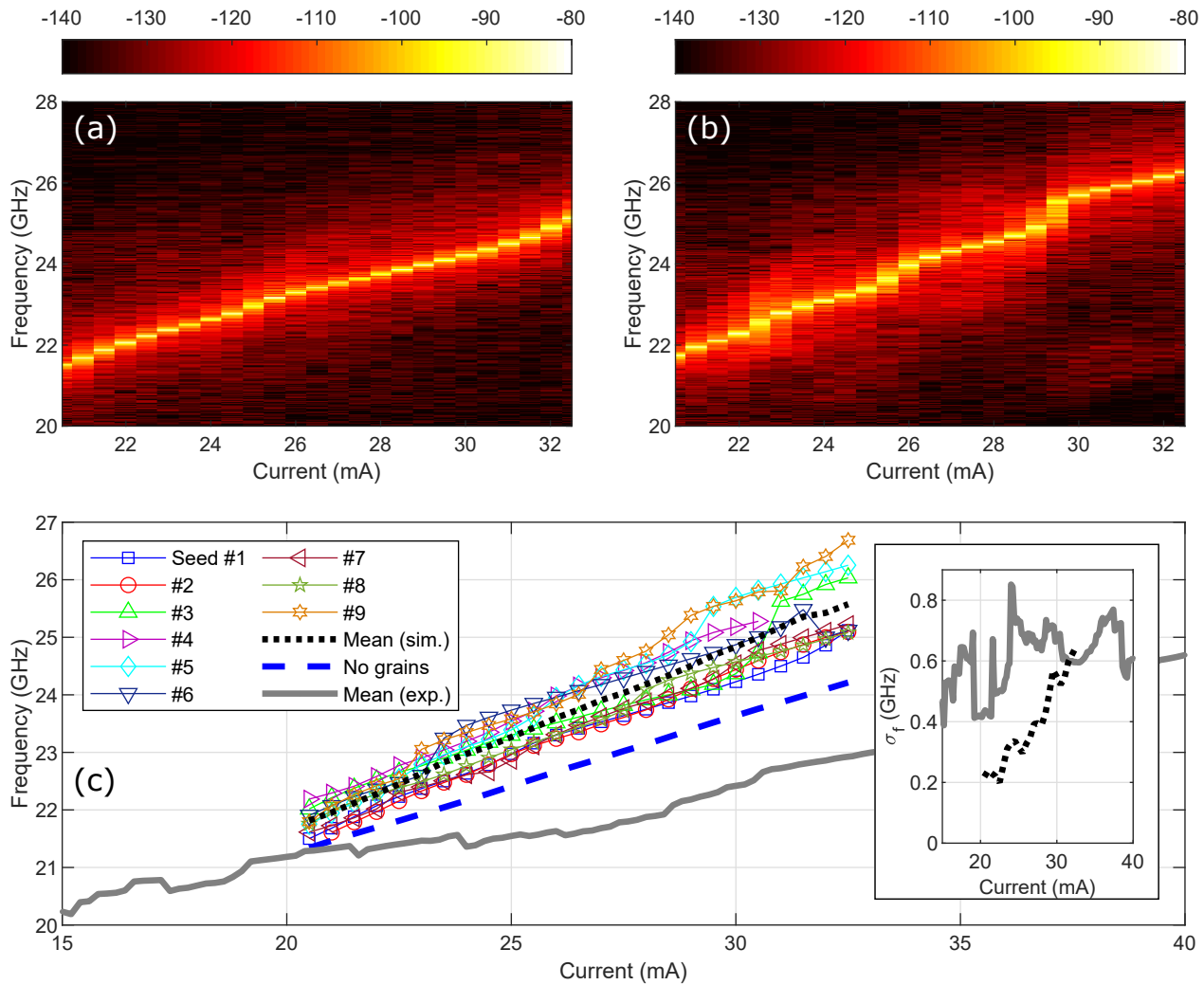}
    \caption[]{Simulations with grain microstructure. Fourier amplitude of $m_y$ (in dB) for: (a) Seed \#1 and (b) Seed \#5. (c) Extracted frequency versus current for nine simulated devices and a homogeneous film. Mean values for simulated and experimental device series. Inset: The standard deviation of the frequency $\sigma_f$ of the simulated and experimental device series.}
    \label{Fig_Sim_Grains}
  \end{center}
\end{figure*}

Figures \ref{Fig_Sim_Grains}(a,b) show the simulated behavior of two grainy free layers with random inter-grain exchange reduction within 0--100 \%. These two "devices" exemplify both continuous and discontinuous nonlinearities: the sample in Figure \ref{Fig_Sim_Grains}(a) show a highly continuous behavior while in Figure \ref{Fig_Sim_Grains}(b) there is a discontinuous frequency transition at 29 mA. The inclusion of grains in the model thus provides a direct explanation of the origin of the $f(I_{\text{DC}})$ nonlinearity in NC-STOs. The grain boundaries form spin wave barriers with varying degree of reflectance and orientation with respect to the propagating spin waves and thus have varying degrees of impact on the STO.

We next investigate the simulated device statistics of $f(I_{\text{DC}})$ for randomly generated grainy films. Figure \ref{Fig_Sim_Grains}(c) shows the extracted frequency versus current for nine simulated devices together with the mean value for both the simulated and experimental (Figure \ref{Fig_NinePSDs}) cases. The homogeneous-film case is included for reference. The simulated grainy films all have their frequencies shifted upwards compared to the nominal homogeneous-film case. The shift of the mean frequency increases gradually with the drive current from 500 MHz at 20.5 mA to 1400 MHz at 32.5 mA. In fact, we do not observe a single frequency below the homogeneous-film case for any simulated device at any current level. This shows that the STO also for the grain microstructure case always selects a standing spin wave pattern that results in an upward rather than downward frequency shift and is in line with the asymmetry towards higher frequency (and amplitude) found in the single-barrier case, Figure \ref{Fig_Sim_Barrier}(a). In the grain case, the standing spin wave pattern can be altered both by changing the number of nodes towards a given grain boundary, or by changing to another dominant grain. The large number of possible mode configurations gives a high probability of always finding a mode with the preferred positive frequency shift.

We note that the experimental mean $f(I_{\text{DC}})$ actually evolves with a factor of two lower slope $df/dI_{\text{DC}}$ than the simulated grain and homogeneous cases. Our free simulation parameters were initially tuned to give largely correct threshold current and frequency for the propagating spin wave mode, but were not adjusted to fit the experimental $f(I_{\text{DC}})$ relation for higher drive currents. We believe that this discrepancy in the simulations might be due to the true spin torque efficiency being lower than the value used for the simulations. It can also be related to the real Oersted field, which is probably lower than what is calculated using the inifite-wire approximation. There is also a lateral spread in current due to the device design \cite{Banuazizi2017}, where the current is intended to flow at the bottom of the spin valve mesa from the NC region out to the ground contacts. The lateral current spread decreases the current density in the NC region and further modifies the Oersted field. We have also not taken into account any possible temperature dependence for the magnetization (i.e., the FMR frequency), which would decrease at higher $I_{\text{DC}}$ due to the higher electrical power dissipation. Exploring this parameter space while fine-tuning the distribution function for the inter-grain exchange coupling to achieve even better correspondence is beyond the scope of the present work. Exact correspondence is also not necessary for discussing the mechanisms of spin wave generation, reflection and interaction and their consequences.

The inset of Figure \ref{Fig_Sim_Grains}(c) shows the standard deviation (between devices) of the frequency $\sigma_f$ for the simulated and experimental devices. The standard deviation of the simulated devices increases linearly over the simulated current interval with a factor of three, from 0.2 GHz to 0.6 GHz. This factor of three coincides with the increase of the grain-induced frequency shift as a function of drive current in Figure \ref{Fig_Sim_Grains}(c). It is a natural behavior that the device-to-device variation that is due to grain-induced frequency shifting is directly proportional to the mean magnitude of the frequency shift.

In the upper half of the simulated current range the standard deviation of the frequency reaches the levels found in the experiment, i.e. $400-600 \textrm{ MHz}$. This quantitative correspondence strengthens the hypothesis of grain-induced spin wave reflection as a main source of the device-to-device variation. Since the trends are not the same (constant versus linearly increasing), there remains modelling aspects in particular for lower drive currents. The most straight-forward approach would be to further decrease the inter-grain exchange coupling to force the grain effect down to lower oscillation amplitudes. For the experimental devices at lower drive current, there is also a more prominent appearance of the pre-threshold mode which perturbs the propagating mode in the devices in Figure \ref{Fig_NinePSDs}(d,e,h,i). It is beyond the scope of this paper to fully reproduce the pre-threshold behavior.

\begin{figure}[t]
  \begin{center}
	\includegraphics[trim = 0mm 0mm 0mm 0mm, clip]{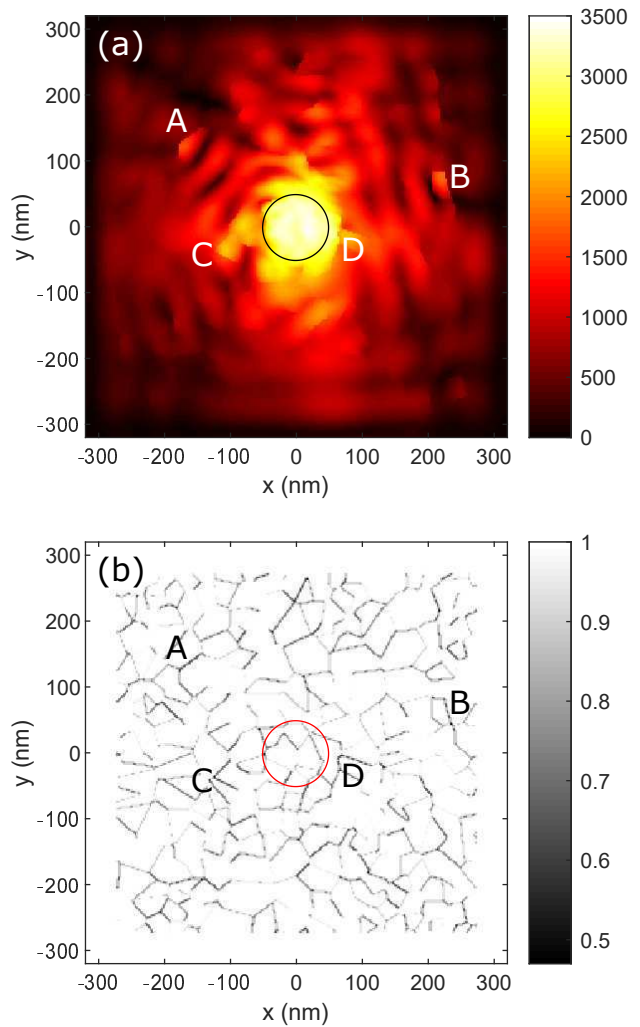}
    \caption[]{Simulations of the grain structure with seed \#5. (a) Fourier amplitude of $m_y$ for the propagating mode at its peak frequency 22.7 GHz for $I_{\text{DC}} = \textrm{23.5 mA}$. (b) The underlying grain structure, showing the normalized strength of the inter-grain exchange coupling. The nano-contact region is indicated by the circle.}
    \label{Fig_Sim_GrainOscPattern}
  \end{center}
\end{figure}

Figure \ref{Fig_Sim_GrainOscPattern}(a) shows the mode structure for seed \#5 at $I_{\text{DC}} = \textrm{23.5 mA}$. The structure is shown as the time-averaged oscillation amplitude, as opposed to an instantaneous snapshot of the propagating waves. The oscillation amplitude falls off non-monotonously outside the nano-contact and forms nodes and anti-nodes in a complex interference pattern. Points of spin wave reflection can be identified where there is a discontinuous drop in the oscillation amplitude. Four clear reflectors are indicated at points A, B, C and D in Figure \ref{Fig_Sim_GrainOscPattern}(a). Looking at the exchange coupling at the same points in Figure \ref{Fig_Sim_GrainOscPattern}(b) reveals that the reflection occurs at grain boundaries where the exchange coupling has been strongly reduced. Grain boundaries A and B are oriented so that their normal direction is pointing approximately towards the NC. At point C there are two possible strong reflectors oriented at approximately $45^{\circ}$ angle relative to the NC direction and it is not clear that they reflect spin waves back directly to the NC. However, the proximity to the NC still results in interference effects at strong amplitudes close to the main oscillation. Around point D there are multiple grain boundaries that are too close to the NC to form a node and anti-node, but that nonetheless act to confine the oscillation. The STO stabilizes at the stationary oscillation state that results in the least amount of conflict between the reflected spin waves returning from points A, B, C and D.

\begin{figure}[t]
  \begin{center}
	\includegraphics[trim = 0mm 0mm 0mm 0mm, clip]{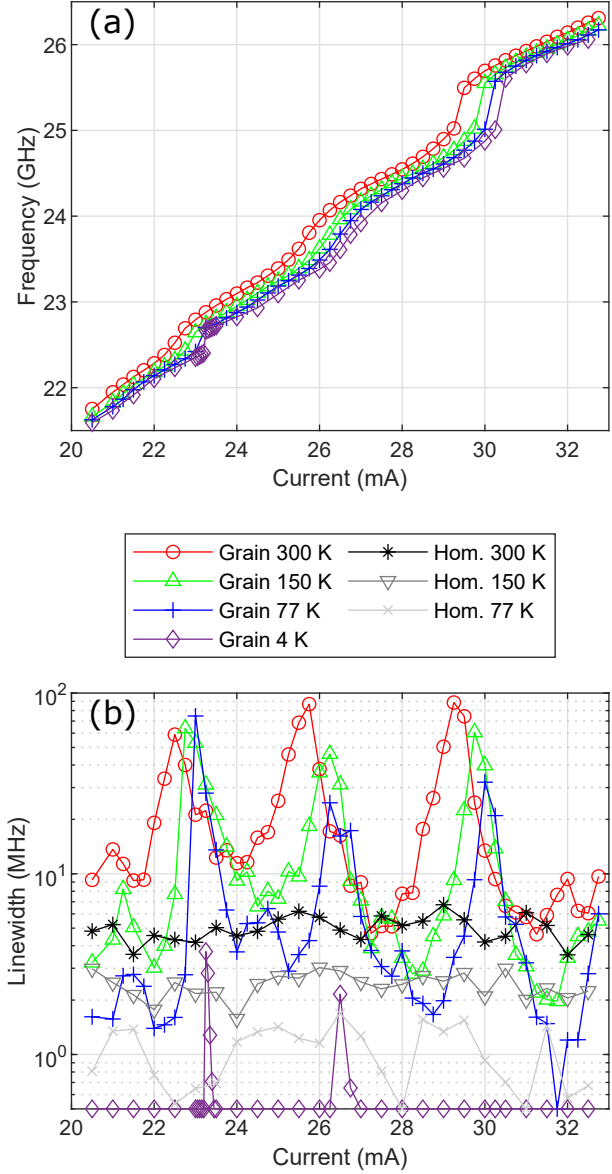}
    \caption[]{Simulation of the seed \#5 grain microstructure and the homogeneous film case at sample temperatures $T$ = 300, 150, 77 and 4 K. (a) Frequency versus drive current. (b) Spectral linewidth (FWHM) versus drive current. Linewidth values below our effective resolution of 0.5 MHz have been plotted at this value. The homogeneous film at $T = 4\textrm{ K}$ is below the linewidth resolution for the entire current range and has been omitted.}
    \label{Fig_Sim_Linewidth}
  \end{center}
\end{figure}

We finally investigate the impact of the grain structure on the frequency stability of the oscillation. For this study we select seed \#5 since it contains the cases of both continuous and discontinuous nonlinearity in $f(I_{\text{DC}})$ and compare its spectral linewidth to the homogeneous film case. Simulations were carried out for sample temperatures $T$ of 300, 150, 77 and 4 K. Figure \ref{Fig_Sim_Linewidth} shows the frequency and spectral linewidth (full width at half maximum) as functions of the drive current $I_{\text{DC}}$.

Figure \ref{Fig_Sim_Linewidth}(a) shows that as the temperature is decreased, the frequency for a given current also decreases. This simulation result agrees with our previous experimental results \cite{Muduli2012b} for similar devices and magnetic field. More interestingly, the temperature decrease induces a shift of the nonlinear operating points to higher drive currents. The nonlinearity is primarily a function of the frequency, which can best be seen at the operating point around 30 mA, where there is a discontinuous step which goes from 25 GHz up to 25.5 GHz for all temperatures. This is a natural consequence of the standing spin wave landscape, where the operating frequency is set by the optimum spin wave length. The nonlinearity around 26 mA is continuous for all temperatures.

The nonlinearity at 23 mA changes character as the temperature changes: at 300 K it is continuous but breaks up into a discontinuous transition at 150, 77 and 4 K. This illustrates that there is no fundamental difference in the origin of the continuous and discontinuous nonlinearity. In the discontinuous case there are two resonance states where the oscillator selects one at a time. Around the transition, thermal energy may be able to kick the trajectory back and forth between the states which can be observed as mode-jumping \cite{Eklund2013,Malm2019}. In the continuous case, the oscillator enters a trajectory that is intermediate to the two underlying resonances. In this continuous transition case, the nano-contact spin torque oscillator can readily be analyzed within the framework of the general nonlinear auto-oscillator theory.

Figure \ref{Fig_Sim_Linewidth}(b) shows that the nonlinear operating points are associated with a destabilization of the frequency. For homogenous films, the linewidth is largely independent of the drive current and reaches a maximum value of 6.7 MHz for $T = 300 \textrm{ K}$. With the grain microstructure, the linewith varies from the homogenous-film values of single MHz inside the linear regions up to 85-90 MHz at the nonlinear operating points. As the temperature is decreased, the maximum linewidth points shift to higher currents. This occurs since the entire $f(I_{\text{DC}})$ relation is moved to higher currents, as previously discussed for Figure \ref{Fig_Sim_Linewidth}(a). The two nonlinearities at higher current (26 and 30 mA) both show a decreasing value for the maximum linewidth as the temperature is decreased. This is the generally expected behavior for single modes described by the nonlinear auto-oscillator theory\cite{Slavin2009a}, where the nonlinar amplification factor $\nu$ along with the temperature determines the linewidth. The same nonlinear amplification factor has also been shown to be applicable in the multi-mode case\cite{Heinonen2013} with thermally activated mode-jumping. The situation of mode-hopping between multiple excitable modes creates an increased sensitivity to thermal fluctuations of the oscillation power through a decrease in the power restoration rate $\Gamma_p$. Since $\nu \sim 1/\Gamma_p$, this theory can be used to at least qualitatively explain the substantially increased linewidth at the nonlinear operating points. Conversely, our work explains the origin of the different modes and gives a physical justification of the applicability of the multi-mode theory for the analysis of the nano-contact STO as carried out in Ref. \citenum{Heinonen2013}.

The low-current nonlinearity at 23 mA again shows a different behavior. Here the maximum linewidth instead increases when the temperature is decreased from 300 K to 77 K. At 4 K the linewidth has decreased to single-MHz values, similar to the case of a well-defined discontinuous nonlinearity. Since this nonlinearity changes its nature from continuous to discontinuous when changing $T$ from 300 K to 77 K we cannot expect neither single- nor multi-mode theory to accurately describe the temperature dependence of the frequency stability across the transition. There may also be additional instability induced by the simultaneous availability of both single- and multi-mode solutions. More detailed study of the transition from continuous into discontinuous nonlinearity is beyond the scope of this work.

\section{Conclusions}
Experimental spectra from nine nominally identical devices has been presented and their qualitative behavior have been described. The sample-to-sample variation in terms of the frequency as a function of current is significant quantitatively as well as qualitatively, with a common feature being linear regions that are connected by nonlinearities that can be either continuous or discontinuous (in the form of a frequency step). This qualitative behavior has been reproduced in simulations incorporating the $\sim$30 nm grain structure measured using AFM and SEM, with randomly reduced inter-grain exchange coupling. The reduction of the inter-grain exchange coupling results in spin wave reflection, which in turn facilitates self-locking of the STO to geometry-defined resonant frequencies. The spin wave resonance preferably acts to increase the oscillation frequency compared to the homogeneous film case. Each of the strongly reflecting grain boundaries constitutes one resonance condition and the final frequency selection for a given current is determined by the inherent STO frequency and the relative strengths of the different reflections. The different standing spin wave modes act to increase the spectral linewidth by more than one order of magnitude at operating points where several of them are simultaneously excitable.

In all, the investigation shows that spin wave reflection and resonance against grain boundaries constitute a physically reasonable model that is able to explain the origin of the continuous and discontinuous nonlinearities in the frequency versus current. This model also explains a large part of the device-to-device variation as stemming from the random grain structure, with partially quantitative agreement with the experimental device variation. For improved agreement, we suggest future modelling work to further reduce the inter-grain exchange coupling and increase the effects of spin wave reflection to its maximum. Other possible sources of variability are inhomogeneity in the magnetic parameters such as the saturation magnetization (or film thickness) and spin polarization ratio. Given the grain microstructure and assuming columnar growth throughout the thin film stack, it would be natural to assign these varying properties at the grain level.

The different grain-induced resonance conditions can be viewed as separate spin wave sub-modes which are simultaneously excitable at the nonlinear operating points. This instability explains the elevated level of the spectral linewidth at the nonlinearities, provides a physical motivation for multi-mode oscillator theory and explains the origin of the apparent nonlinearity as it is treated in single-mode theory. Since improving the crystalline quality of sputtered metallic films is considered to be difficult, further work on controlling and improving the device performance is suggested to focus also on the machining of artificial spin wave reflection boundaries that can dominate over the random grain boundaries.

\begin{acknowledgments}
The authors gratefully acknowledge Ahmad Abedin for assistance with the SEM measurements and Federico Pever\'e and Faraz Khavari for assistance with the AFM measurements. We thank the PDC Center for High Performance Computing, KTH Royal Institute of Technology, Sweden, for providing access to part of the computing resources used in this research. The computational results presented have in part been achieved using the Vienna Scientific Cluster (VSC). This work was financially supported by the Swedish Research Council (VR) through the projects 2017-04196 and 2012-05372.
\end{acknowledgments}

\bibliography{references}

\begin{thebibliography}{51}%
\makeatletter
\providecommand \@ifxundefined [1]{%
 \@ifx{#1\undefined}
}%
\providecommand \@ifnum [1]{%
 \ifnum #1\expandafter \@firstoftwo
 \else \expandafter \@secondoftwo
 \fi
}%
\providecommand \@ifx [1]{%
 \ifx #1\expandafter \@firstoftwo
 \else \expandafter \@secondoftwo
 \fi
}%
\providecommand \natexlab [1]{#1}%
\providecommand \enquote  [1]{``#1''}%
\providecommand \bibnamefont  [1]{#1}%
\providecommand \bibfnamefont [1]{#1}%
\providecommand \citenamefont [1]{#1}%
\providecommand \href@noop [0]{\@secondoftwo}%
\providecommand \href [0]{\begingroup \@sanitize@url \@href}%
\providecommand \@href[1]{\@@startlink{#1}\@@href}%
\providecommand \@@href[1]{\endgroup#1\@@endlink}%
\providecommand \@sanitize@url [0]{\catcode `\\12\catcode `\$12\catcode
  `\&12\catcode `\#12\catcode `\^12\catcode `\_12\catcode `\%12\relax}%
\providecommand \@@startlink[1]{}%
\providecommand \@@endlink[0]{}%
\providecommand \url  [0]{\begingroup\@sanitize@url \@url }%
\providecommand \@url [1]{\endgroup\@href {#1}{\urlprefix }}%
\providecommand \urlprefix  [0]{URL }%
\providecommand \Eprint [0]{\href }%
\providecommand \doibase [0]{http://dx.doi.org/}%
\providecommand \selectlanguage [0]{\@gobble}%
\providecommand \bibinfo  [0]{\@secondoftwo}%
\providecommand \bibfield  [0]{\@secondoftwo}%
\providecommand \translation [1]{[#1]}%
\providecommand \BibitemOpen [0]{}%
\providecommand \bibitemStop [0]{}%
\providecommand \bibitemNoStop [0]{.\EOS\space}%
\providecommand \EOS [0]{\spacefactor3000\relax}%
\providecommand \BibitemShut  [1]{\csname bibitem#1\endcsname}%
\let\auto@bib@innerbib\@empty
\bibitem [{\citenamefont {Silva}\ and\ \citenamefont
  {Rippard}(2008)}]{Silva2008}%
  \BibitemOpen
  \bibfield  {author} {\bibinfo {author} {\bibfnamefont {T.~J.}\ \bibnamefont
  {Silva}}\ and\ \bibinfo {author} {\bibfnamefont {W.~H.}\ \bibnamefont
  {Rippard}},\ }\href {\doibase http://dx.doi.org/10.1016/j.jmmm.2007.12.022}
  {\bibfield  {journal} {\bibinfo  {journal} {J. Magn. Magn. Mater.}\ }\textbf
  {\bibinfo {volume} {320}},\ \bibinfo {pages} {1260 } (\bibinfo {year}
  {2008})}\BibitemShut {NoStop}%
\bibitem [{\citenamefont {{Chen}}\ \emph {et~al.}(2016)\citenamefont {{Chen}},
  \citenamefont {{Dumas}}, \citenamefont {{Eklund}}, \citenamefont {{Muduli}},
  \citenamefont {{Houshang}}, \citenamefont {{Awad}}, \citenamefont
  {{D\"urrenfeld}}, \citenamefont {{Malm}}, \citenamefont {{Rusu}},\ and\
  \citenamefont {{{\AA}kerman}}}]{Chen2016}%
  \BibitemOpen
  \bibfield  {author} {\bibinfo {author} {\bibfnamefont {T.}~\bibnamefont
  {{Chen}}}, \bibinfo {author} {\bibfnamefont {R.~K.}\ \bibnamefont {{Dumas}}},
  \bibinfo {author} {\bibfnamefont {A.}~\bibnamefont {{Eklund}}}, \bibinfo
  {author} {\bibfnamefont {P.~K.}\ \bibnamefont {{Muduli}}}, \bibinfo {author}
  {\bibfnamefont {A.}~\bibnamefont {{Houshang}}}, \bibinfo {author}
  {\bibfnamefont {A.~A.}\ \bibnamefont {{Awad}}}, \bibinfo {author}
  {\bibfnamefont {P.}~\bibnamefont {{D\"urrenfeld}}}, \bibinfo {author}
  {\bibfnamefont {B.~G.}\ \bibnamefont {{Malm}}}, \bibinfo {author}
  {\bibfnamefont {A.}~\bibnamefont {{Rusu}}}, \ and\ \bibinfo {author}
  {\bibfnamefont {J.}~\bibnamefont {{{\AA}kerman}}},\ }\href {\doibase
  10.1109/JPROC.2016.2554518} {\bibfield  {journal} {\bibinfo  {journal}
  {Proceedings of the IEEE}\ }\textbf {\bibinfo {volume} {104}},\ \bibinfo
  {pages} {1919} (\bibinfo {year} {2016})}\BibitemShut {NoStop}%
\bibitem [{\citenamefont {Slonczewski}(1996)}]{Slonczewski1996a}%
  \BibitemOpen
  \bibfield  {author} {\bibinfo {author} {\bibfnamefont {J.}~\bibnamefont
  {Slonczewski}},\ }\href {\doibase 10.1016/0304-8853(96)00062-5} {\bibfield
  {journal} {\bibinfo  {journal} {J. Magn. Magn. Mater.}\ }\textbf {\bibinfo
  {volume} {159}},\ \bibinfo {pages} {L1} (\bibinfo {year} {1996})}\BibitemShut
  {NoStop}%
\bibitem [{\citenamefont {Berger}(1996)}]{Berger1996}%
  \BibitemOpen
  \bibfield  {author} {\bibinfo {author} {\bibfnamefont {L.}~\bibnamefont
  {Berger}},\ }\href {10.1103/PhysRevB.54.9353} {\bibfield  {journal} {\bibinfo
   {journal} {Phys. Rev. B}\ }\textbf {\bibinfo {volume} {54}},\ \bibinfo
  {pages} {9353} (\bibinfo {year} {1996})}\BibitemShut {NoStop}%
\bibitem [{\citenamefont {Ralph}\ and\ \citenamefont
  {Stiles}(2008)}]{Ralph2008a}%
  \BibitemOpen
  \bibfield  {author} {\bibinfo {author} {\bibfnamefont {D.}~\bibnamefont
  {Ralph}}\ and\ \bibinfo {author} {\bibfnamefont {M.}~\bibnamefont {Stiles}},\
  }\href {\doibase 10.1016/j.jmmm.2007.12.019} {\bibfield  {journal} {\bibinfo
  {journal} {J. Magn. Magn. Mater.}\ }\textbf {\bibinfo {volume} {320}},\
  \bibinfo {pages} {1190} (\bibinfo {year} {2008})}\BibitemShut {NoStop}%
\bibitem [{\citenamefont {Baibich}\ \emph {et~al.}(1988)\citenamefont
  {Baibich}, \citenamefont {Broto}, \citenamefont {Fert}, \citenamefont {{Van
  Dau}}, \citenamefont {Petroff}, \citenamefont {Etienne}, \citenamefont
  {Creuzet}, \citenamefont {Friederich},\ and\ \citenamefont
  {Chazelas}}]{Baibich1988}%
  \BibitemOpen
  \bibfield  {author} {\bibinfo {author} {\bibfnamefont {M.~N.}\ \bibnamefont
  {Baibich}}, \bibinfo {author} {\bibfnamefont {J.~M.}\ \bibnamefont {Broto}},
  \bibinfo {author} {\bibfnamefont {A.}~\bibnamefont {Fert}}, \bibinfo {author}
  {\bibfnamefont {F.~N.}\ \bibnamefont {{Van Dau}}}, \bibinfo {author}
  {\bibfnamefont {F.}~\bibnamefont {Petroff}}, \bibinfo {author} {\bibfnamefont
  {P.}~\bibnamefont {Etienne}}, \bibinfo {author} {\bibfnamefont
  {G.}~\bibnamefont {Creuzet}}, \bibinfo {author} {\bibfnamefont
  {A.}~\bibnamefont {Friederich}}, \ and\ \bibinfo {author} {\bibfnamefont
  {J.}~\bibnamefont {Chazelas}},\ }\href {\doibase 10.1103/PhysRevLett.61.2472}
  {\bibfield  {journal} {\bibinfo  {journal} {Phys. Rev. Lett.}\ }\textbf
  {\bibinfo {volume} {61}},\ \bibinfo {pages} {2472} (\bibinfo {year}
  {1988})}\BibitemShut {NoStop}%
\bibitem [{\citenamefont {Binasch}\ \emph {et~al.}(1989)\citenamefont
  {Binasch}, \citenamefont {Gr\"{u}nberg}, \citenamefont {Saurenbach},\ and\
  \citenamefont {Zinn}}]{Binasch1989}%
  \BibitemOpen
  \bibfield  {author} {\bibinfo {author} {\bibfnamefont {G.}~\bibnamefont
  {Binasch}}, \bibinfo {author} {\bibfnamefont {P.}~\bibnamefont
  {Gr\"{u}nberg}}, \bibinfo {author} {\bibfnamefont {F.}~\bibnamefont
  {Saurenbach}}, \ and\ \bibinfo {author} {\bibfnamefont {W.}~\bibnamefont
  {Zinn}},\ }\href {\doibase 10.1103/PhysRevB.39.4828} {\bibfield  {journal}
  {\bibinfo  {journal} {Phys. Rev. B}\ }\textbf {\bibinfo {volume} {39}},\
  \bibinfo {pages} {4828} (\bibinfo {year} {1989})}\BibitemShut {NoStop}%
\bibitem [{\citenamefont {Pufall}\ \emph {et~al.}(2007)\citenamefont {Pufall},
  \citenamefont {Rippard}, \citenamefont {Schneider},\ and\ \citenamefont
  {Russek}}]{Pufall2007}%
  \BibitemOpen
  \bibfield  {author} {\bibinfo {author} {\bibfnamefont {M.~R.}\ \bibnamefont
  {Pufall}}, \bibinfo {author} {\bibfnamefont {W.~H.}\ \bibnamefont {Rippard}},
  \bibinfo {author} {\bibfnamefont {M.~L.}\ \bibnamefont {Schneider}}, \ and\
  \bibinfo {author} {\bibfnamefont {S.~E.}\ \bibnamefont {Russek}},\ }\href
  {\doibase 10.1103/PhysRevB.75.140404} {\bibfield  {journal} {\bibinfo
  {journal} {Phys. Rev. B}\ }\textbf {\bibinfo {volume} {75}},\ \bibinfo
  {pages} {140404} (\bibinfo {year} {2007})}\BibitemShut {NoStop}%
\bibitem [{\citenamefont {Mistral}\ \emph {et~al.}(2008)\citenamefont
  {Mistral}, \citenamefont {van Kampen}, \citenamefont {Hrkac}, \citenamefont
  {Kim}, \citenamefont {Devolder}, \citenamefont {Crozat}, \citenamefont
  {Chappert}, \citenamefont {Lagae},\ and\ \citenamefont
  {Schrefl}}]{Mistral2008}%
  \BibitemOpen
  \bibfield  {author} {\bibinfo {author} {\bibfnamefont {Q.}~\bibnamefont
  {Mistral}}, \bibinfo {author} {\bibfnamefont {M.}~\bibnamefont {van Kampen}},
  \bibinfo {author} {\bibfnamefont {G.}~\bibnamefont {Hrkac}}, \bibinfo
  {author} {\bibfnamefont {J.-V.}\ \bibnamefont {Kim}}, \bibinfo {author}
  {\bibfnamefont {T.}~\bibnamefont {Devolder}}, \bibinfo {author}
  {\bibfnamefont {P.}~\bibnamefont {Crozat}}, \bibinfo {author} {\bibfnamefont
  {C.}~\bibnamefont {Chappert}}, \bibinfo {author} {\bibfnamefont
  {L.}~\bibnamefont {Lagae}}, \ and\ \bibinfo {author} {\bibfnamefont
  {T.}~\bibnamefont {Schrefl}},\ }\href {\doibase
  10.1103/PhysRevLett.100.257201} {\bibfield  {journal} {\bibinfo  {journal}
  {Phys. Rev. Lett.}\ }\textbf {\bibinfo {volume} {100}},\ \bibinfo {pages}
  {257201} (\bibinfo {year} {2008})}\BibitemShut {NoStop}%
\bibitem [{\citenamefont {Bonetti}\ \emph {et~al.}(2009)\citenamefont
  {Bonetti}, \citenamefont {Muduli}, \citenamefont {Mancoff},\ and\
  \citenamefont {{\AA}kerman}}]{Bonetti2009}%
  \BibitemOpen
  \bibfield  {author} {\bibinfo {author} {\bibfnamefont {S.}~\bibnamefont
  {Bonetti}}, \bibinfo {author} {\bibfnamefont {P.}~\bibnamefont {Muduli}},
  \bibinfo {author} {\bibfnamefont {F.}~\bibnamefont {Mancoff}}, \ and\
  \bibinfo {author} {\bibfnamefont {J.}~\bibnamefont {{\AA}kerman}},\ }\href
  {\doibase 10.1063/1.3097238} {\bibfield  {journal} {\bibinfo  {journal}
  {Appl. Phys. Lett.}\ }\textbf {\bibinfo {volume} {94}},\ \bibinfo {pages}
  {102507} (\bibinfo {year} {2009})}\BibitemShut {NoStop}%
\bibitem [{\citenamefont {Slavin}\ and\ \citenamefont
  {Tiberkevich}(2005)}]{Slavin2005}%
  \BibitemOpen
  \bibfield  {author} {\bibinfo {author} {\bibfnamefont {A.}~\bibnamefont
  {Slavin}}\ and\ \bibinfo {author} {\bibfnamefont {V.}~\bibnamefont
  {Tiberkevich}},\ }\href {\doibase 10.1103/PhysRevLett.95.237201} {\bibfield
  {journal} {\bibinfo  {journal} {Phys. Rev. Lett.}\ }\textbf {\bibinfo
  {volume} {95}},\ \bibinfo {pages} {237201} (\bibinfo {year}
  {2005})}\BibitemShut {NoStop}%
\bibitem [{\citenamefont {Slonczewski}(1999)}]{Slonczewski1999a}%
  \BibitemOpen
  \bibfield  {author} {\bibinfo {author} {\bibfnamefont {J.}~\bibnamefont
  {Slonczewski}},\ }\href {\doibase 10.1016/S0304-8853(99)00043-8} {\bibfield
  {journal} {\bibinfo  {journal} {J. Magn. Magn. Mater.}\ }\textbf {\bibinfo
  {volume} {195}},\ \bibinfo {pages} {261} (\bibinfo {year}
  {1999})}\BibitemShut {NoStop}%
\bibitem [{\citenamefont {Madami}\ \emph {et~al.}(2011)\citenamefont {Madami},
  \citenamefont {Bonetti}, \citenamefont {Consolo}, \citenamefont {Tacchi},
  \citenamefont {Carlotti}, \citenamefont {Gubbiotti}, \citenamefont {Mancoff},
  \citenamefont {Yar},\ and\ \citenamefont {\AA{}kerman}}]{Madami2011}%
  \BibitemOpen
  \bibfield  {author} {\bibinfo {author} {\bibfnamefont {M.}~\bibnamefont
  {Madami}}, \bibinfo {author} {\bibfnamefont {S.}~\bibnamefont {Bonetti}},
  \bibinfo {author} {\bibfnamefont {G.}~\bibnamefont {Consolo}}, \bibinfo
  {author} {\bibfnamefont {S.}~\bibnamefont {Tacchi}}, \bibinfo {author}
  {\bibfnamefont {G.}~\bibnamefont {Carlotti}}, \bibinfo {author}
  {\bibfnamefont {G.}~\bibnamefont {Gubbiotti}}, \bibinfo {author}
  {\bibfnamefont {F.~B.}\ \bibnamefont {Mancoff}}, \bibinfo {author}
  {\bibfnamefont {M.~A.}\ \bibnamefont {Yar}}, \ and\ \bibinfo {author}
  {\bibfnamefont {J.}~\bibnamefont {\AA{}kerman}},\ }\href {\doibase
  10.1038/nnano.2011.140} {\bibfield  {journal} {\bibinfo  {journal} {Nature
  Nanotech.}\ }\textbf {\bibinfo {volume} {6}},\ \bibinfo {pages} {635}
  (\bibinfo {year} {2011})}\BibitemShut {NoStop}%
\bibitem [{\citenamefont {Madami}\ \emph {et~al.}(2015)\citenamefont {Madami},
  \citenamefont {Iacocca}, \citenamefont {Sani}, \citenamefont {Gubbiotti},
  \citenamefont {Tacchi}, \citenamefont {Dumas}, \citenamefont {\AA{}kerman},\
  and\ \citenamefont {Carlotti}}]{Madami2015}%
  \BibitemOpen
  \bibfield  {author} {\bibinfo {author} {\bibfnamefont {M.}~\bibnamefont
  {Madami}}, \bibinfo {author} {\bibfnamefont {E.}~\bibnamefont {Iacocca}},
  \bibinfo {author} {\bibfnamefont {S.}~\bibnamefont {Sani}}, \bibinfo {author}
  {\bibfnamefont {G.}~\bibnamefont {Gubbiotti}}, \bibinfo {author}
  {\bibfnamefont {S.}~\bibnamefont {Tacchi}}, \bibinfo {author} {\bibfnamefont
  {R.~K.}\ \bibnamefont {Dumas}}, \bibinfo {author} {\bibfnamefont
  {J.}~\bibnamefont {\AA{}kerman}}, \ and\ \bibinfo {author} {\bibfnamefont
  {G.}~\bibnamefont {Carlotti}},\ }\href {\doibase 10.1103/PhysRevB.92.024403}
  {\bibfield  {journal} {\bibinfo  {journal} {Phys. Rev. B}\ }\textbf {\bibinfo
  {volume} {92}},\ \bibinfo {pages} {024403} (\bibinfo {year}
  {2015})}\BibitemShut {NoStop}%
\bibitem [{\citenamefont {Bonetti}\ \emph {et~al.}(2010)\citenamefont
  {Bonetti}, \citenamefont {Tiberkevich}, \citenamefont {Consolo},
  \citenamefont {Finocchio}, \citenamefont {Muduli}, \citenamefont {Mancoff},
  \citenamefont {Slavin},\ and\ \citenamefont {\AA{}kerman}}]{Bonetti2010}%
  \BibitemOpen
  \bibfield  {author} {\bibinfo {author} {\bibfnamefont {S.}~\bibnamefont
  {Bonetti}}, \bibinfo {author} {\bibfnamefont {V.}~\bibnamefont
  {Tiberkevich}}, \bibinfo {author} {\bibfnamefont {G.}~\bibnamefont
  {Consolo}}, \bibinfo {author} {\bibfnamefont {G.}~\bibnamefont {Finocchio}},
  \bibinfo {author} {\bibfnamefont {P.}~\bibnamefont {Muduli}}, \bibinfo
  {author} {\bibfnamefont {F.}~\bibnamefont {Mancoff}}, \bibinfo {author}
  {\bibfnamefont {A.}~\bibnamefont {Slavin}}, \ and\ \bibinfo {author}
  {\bibfnamefont {J.}~\bibnamefont {\AA{}kerman}},\ }\href {\doibase
  10.1103/PhysRevLett.105.217204} {\bibfield  {journal} {\bibinfo  {journal}
  {Phys. Rev. Lett.}\ }\textbf {\bibinfo {volume} {105}},\ \bibinfo {pages}
  {217204} (\bibinfo {year} {2010})}\BibitemShut {NoStop}%
\bibitem [{\citenamefont {Kruglyak}\ \emph {et~al.}(2010)\citenamefont
  {Kruglyak}, \citenamefont {Demokritov},\ and\ \citenamefont
  {Grundler}}]{Kruglyak2010}%
  \BibitemOpen
  \bibfield  {author} {\bibinfo {author} {\bibfnamefont {V.~V.}\ \bibnamefont
  {Kruglyak}}, \bibinfo {author} {\bibfnamefont {S.~O.}\ \bibnamefont
  {Demokritov}}, \ and\ \bibinfo {author} {\bibfnamefont {D.}~\bibnamefont
  {Grundler}},\ }\href {\doibase 10.1088/0022-3727/43/26/260301} {\bibfield
  {journal} {\bibinfo  {journal} {J. Phys. D: Appl. Phys.}\ }\textbf {\bibinfo
  {volume} {43}},\ \bibinfo {pages} {260301} (\bibinfo {year}
  {2010})}\BibitemShut {NoStop}%
\bibitem [{\citenamefont {Bonetti}\ and\ \citenamefont
  {{\AA}kerman}(2013)}]{Bonetti2013}%
  \BibitemOpen
  \bibfield  {author} {\bibinfo {author} {\bibfnamefont {S.}~\bibnamefont
  {Bonetti}}\ and\ \bibinfo {author} {\bibfnamefont {J.}~\bibnamefont
  {{\AA}kerman}},\ }\enquote {\bibinfo {title} {Nano-contact spin-torque
  oscillators as magnonic building blocks},}\ in\ \href {\doibase
  10.1007/978-3-642-30247-3_13} {\emph {\bibinfo {booktitle} {Magnonics: From
  Fundamentals to Applications}}},\ \bibinfo {editor} {edited by\ \bibinfo
  {editor} {\bibfnamefont {S.~O.}\ \bibnamefont {Demokritov}}\ and\ \bibinfo
  {editor} {\bibfnamefont {A.~N.}\ \bibnamefont {Slavin}}}\ (\bibinfo
  {publisher} {Springer Berlin Heidelberg},\ \bibinfo {address} {Berlin,
  Heidelberg},\ \bibinfo {year} {2013})\ pp.\ \bibinfo {pages}
  {177--187}\BibitemShut {NoStop}%
\bibitem [{\citenamefont {Chumak}\ \emph {et~al.}(2015)\citenamefont {Chumak},
  \citenamefont {Vasyuchka}, \citenamefont {Serga},\ and\ \citenamefont
  {Hillebrands}}]{Chumak2015}%
  \BibitemOpen
  \bibfield  {author} {\bibinfo {author} {\bibfnamefont {A.~V.}\ \bibnamefont
  {Chumak}}, \bibinfo {author} {\bibfnamefont {V.~I.}\ \bibnamefont
  {Vasyuchka}}, \bibinfo {author} {\bibfnamefont {A.~A.}\ \bibnamefont
  {Serga}}, \ and\ \bibinfo {author} {\bibfnamefont {B.}~\bibnamefont
  {Hillebrands}},\ }\href {\doibase 10.1038/nphys3347} {\bibfield  {journal}
  {\bibinfo  {journal} {Nature Phys.}\ }\textbf {\bibinfo {volume} {11}},\
  \bibinfo {pages} {453} (\bibinfo {year} {2015})}\BibitemShut {NoStop}%
\bibitem [{\citenamefont {Rippard}\ \emph {et~al.}(2006)\citenamefont
  {Rippard}, \citenamefont {Pufall},\ and\ \citenamefont
  {Russek}}]{Rippard2006}%
  \BibitemOpen
  \bibfield  {author} {\bibinfo {author} {\bibfnamefont {W.~H.}\ \bibnamefont
  {Rippard}}, \bibinfo {author} {\bibfnamefont {M.~R.}\ \bibnamefont {Pufall}},
  \ and\ \bibinfo {author} {\bibfnamefont {S.~E.}\ \bibnamefont {Russek}},\
  }\href {\doibase 10.1103/PhysRevB.74.224409} {\bibfield  {journal} {\bibinfo
  {journal} {Phys. Rev. B}\ }\textbf {\bibinfo {volume} {74}},\ \bibinfo
  {pages} {224409} (\bibinfo {year} {2006})}\BibitemShut {NoStop}%
\bibitem [{\citenamefont {Tamaru}\ and\ \citenamefont
  {Ricketts}(2012)}]{Tamaru2012}%
  \BibitemOpen
  \bibfield  {author} {\bibinfo {author} {\bibfnamefont {S.}~\bibnamefont
  {Tamaru}}\ and\ \bibinfo {author} {\bibfnamefont {D.~S.}\ \bibnamefont
  {Ricketts}},\ }\href {\doibase 10.1109/LMAG.2012.2208098} {\bibfield
  {journal} {\bibinfo  {journal} {IEEE Magn. Lett.}\ }\textbf {\bibinfo
  {volume} {3}},\ \bibinfo {pages} {3000504} (\bibinfo {year}
  {2012})}\BibitemShut {NoStop}%
\bibitem [{\citenamefont {Houshang}\ \emph {et~al.}(2018)\citenamefont
  {Houshang}, \citenamefont {Khymyn}, \citenamefont {Fulara}, \citenamefont
  {Gangwar}, \citenamefont {Haidar}, \citenamefont {Etesami}, \citenamefont
  {Ferreira}, \citenamefont {Freitas}, \citenamefont {Dvornik}, \citenamefont
  {Dumas},\ and\ \citenamefont {{\AA}kerman}}]{Houshang2018}%
  \BibitemOpen
  \bibfield  {author} {\bibinfo {author} {\bibfnamefont {A.}~\bibnamefont
  {Houshang}}, \bibinfo {author} {\bibfnamefont {R.}~\bibnamefont {Khymyn}},
  \bibinfo {author} {\bibfnamefont {H.}~\bibnamefont {Fulara}}, \bibinfo
  {author} {\bibfnamefont {A.}~\bibnamefont {Gangwar}}, \bibinfo {author}
  {\bibfnamefont {M.}~\bibnamefont {Haidar}}, \bibinfo {author} {\bibfnamefont
  {S.~R.}\ \bibnamefont {Etesami}}, \bibinfo {author} {\bibfnamefont
  {R.}~\bibnamefont {Ferreira}}, \bibinfo {author} {\bibfnamefont {P.~P.}\
  \bibnamefont {Freitas}}, \bibinfo {author} {\bibfnamefont {M.}~\bibnamefont
  {Dvornik}}, \bibinfo {author} {\bibfnamefont {R.~K.}\ \bibnamefont {Dumas}},
  \ and\ \bibinfo {author} {\bibfnamefont {J.}~\bibnamefont {{\AA}kerman}},\
  }\href {\doibase 10.1038/s41467-018-06589-0} {\bibfield  {journal} {\bibinfo
  {journal} {Nature Communications}\ }\textbf {\bibinfo {volume} {9}},\
  \bibinfo {pages} {4374} (\bibinfo {year} {2018})}\BibitemShut {NoStop}%
\bibitem [{\citenamefont {Slavin}\ and\ \citenamefont
  {Tiberkevich}(2009)}]{Slavin2009a}%
  \BibitemOpen
  \bibfield  {author} {\bibinfo {author} {\bibfnamefont {A.}~\bibnamefont
  {Slavin}}\ and\ \bibinfo {author} {\bibfnamefont {V.}~\bibnamefont
  {Tiberkevich}},\ }\href {\doibase 10.1109/TMAG.2008.2009935} {\bibfield
  {journal} {\bibinfo  {journal} {IEEE Trans. Magn.}\ }\textbf {\bibinfo
  {volume} {45}},\ \bibinfo {pages} {1875} (\bibinfo {year}
  {2009})}\BibitemShut {NoStop}%
\bibitem [{\citenamefont {Eklund}\ \emph {et~al.}(2014)\citenamefont {Eklund},
  \citenamefont {Bonetti}, \citenamefont {Sani}, \citenamefont {Mohseni},
  \citenamefont {Persson}, \citenamefont {Chung}, \citenamefont {Banuazizi},
  \citenamefont {Iacocca}, \citenamefont {\"{O}stling}, \citenamefont
  {{\AA}kerman},\ and\ \citenamefont {Malm}}]{Eklund2014}%
  \BibitemOpen
  \bibfield  {author} {\bibinfo {author} {\bibfnamefont {A.}~\bibnamefont
  {Eklund}}, \bibinfo {author} {\bibfnamefont {S.}~\bibnamefont {Bonetti}},
  \bibinfo {author} {\bibfnamefont {S.~R.}\ \bibnamefont {Sani}}, \bibinfo
  {author} {\bibfnamefont {S.~M.}\ \bibnamefont {Mohseni}}, \bibinfo {author}
  {\bibfnamefont {J.}~\bibnamefont {Persson}}, \bibinfo {author} {\bibfnamefont
  {S.}~\bibnamefont {Chung}}, \bibinfo {author} {\bibfnamefont {S.~A.~H.}\
  \bibnamefont {Banuazizi}}, \bibinfo {author} {\bibfnamefont {E.}~\bibnamefont
  {Iacocca}}, \bibinfo {author} {\bibfnamefont {M.}~\bibnamefont
  {\"{O}stling}}, \bibinfo {author} {\bibfnamefont {J.}~\bibnamefont
  {{\AA}kerman}}, \ and\ \bibinfo {author} {\bibfnamefont {B.~G.}\ \bibnamefont
  {Malm}},\ }\href {\doibase 10.1063/1.4867257} {\bibfield  {journal} {\bibinfo
   {journal} {Appl. Phys. Lett.}\ }\textbf {\bibinfo {volume} {104}},\ \bibinfo
  {pages} {092405} (\bibinfo {year} {2014})}\BibitemShut {NoStop}%
\bibitem [{\citenamefont {Pufall}\ \emph {et~al.}(2012)\citenamefont {Pufall},
  \citenamefont {Rippard}, \citenamefont {Russek},\ and\ \citenamefont
  {Evarts}}]{Pufall2012}%
  \BibitemOpen
  \bibfield  {author} {\bibinfo {author} {\bibfnamefont {M.~R.}\ \bibnamefont
  {Pufall}}, \bibinfo {author} {\bibfnamefont {W.~H.}\ \bibnamefont {Rippard}},
  \bibinfo {author} {\bibfnamefont {S.~E.}\ \bibnamefont {Russek}}, \ and\
  \bibinfo {author} {\bibfnamefont {E.~R.}\ \bibnamefont {Evarts}},\ }\href
  {\doibase 10.1103/PhysRevB.86.094404} {\bibfield  {journal} {\bibinfo
  {journal} {Phys. Rev. B}\ }\textbf {\bibinfo {volume} {86}},\ \bibinfo
  {pages} {094404} (\bibinfo {year} {2012})}\BibitemShut {NoStop}%
\bibitem [{\citenamefont {Hoefer}\ \emph {et~al.}(2008)\citenamefont {Hoefer},
  \citenamefont {Silva},\ and\ \citenamefont {Stiles}}]{Hoefer2008}%
  \BibitemOpen
  \bibfield  {author} {\bibinfo {author} {\bibfnamefont {M.~A.}\ \bibnamefont
  {Hoefer}}, \bibinfo {author} {\bibfnamefont {T.~J.}\ \bibnamefont {Silva}}, \
  and\ \bibinfo {author} {\bibfnamefont {M.~D.}\ \bibnamefont {Stiles}},\
  }\href {\doibase 10.1103/PhysRevB.77.144401} {\bibfield  {journal} {\bibinfo
  {journal} {Phys. Rev. B}\ }\textbf {\bibinfo {volume} {77}},\ \bibinfo
  {pages} {144401} (\bibinfo {year} {2008})}\BibitemShut {NoStop}%
\bibitem [{\citenamefont {Dumas}\ \emph {et~al.}(2013)\citenamefont {Dumas},
  \citenamefont {Iacocca}, \citenamefont {Bonetti}, \citenamefont {Sani},
  \citenamefont {Mohseni}, \citenamefont {Eklund}, \citenamefont {Persson},
  \citenamefont {Heinonen},\ and\ \citenamefont {\AA{}kerman}}]{Dumas2013}%
  \BibitemOpen
  \bibfield  {author} {\bibinfo {author} {\bibfnamefont {R.~K.}\ \bibnamefont
  {Dumas}}, \bibinfo {author} {\bibfnamefont {E.}~\bibnamefont {Iacocca}},
  \bibinfo {author} {\bibfnamefont {S.}~\bibnamefont {Bonetti}}, \bibinfo
  {author} {\bibfnamefont {S.~R.}\ \bibnamefont {Sani}}, \bibinfo {author}
  {\bibfnamefont {S.~M.}\ \bibnamefont {Mohseni}}, \bibinfo {author}
  {\bibfnamefont {A.}~\bibnamefont {Eklund}}, \bibinfo {author} {\bibfnamefont
  {J.}~\bibnamefont {Persson}}, \bibinfo {author} {\bibfnamefont
  {O.}~\bibnamefont {Heinonen}}, \ and\ \bibinfo {author} {\bibfnamefont
  {J.}~\bibnamefont {\AA{}kerman}},\ }\href {\doibase
  10.1103/PhysRevLett.110.257202} {\bibfield  {journal} {\bibinfo  {journal}
  {Phys. Rev. Lett.}\ }\textbf {\bibinfo {volume} {110}},\ \bibinfo {pages}
  {257202} (\bibinfo {year} {2013})}\BibitemShut {NoStop}%
\bibitem [{\citenamefont {Bonetti}\ \emph {et~al.}(2012)\citenamefont
  {Bonetti}, \citenamefont {Puliafito}, \citenamefont {Consolo}, \citenamefont
  {Tiberkevich}, \citenamefont {Slavin},\ and\ \citenamefont
  {\AA{}kerman}}]{Bonetti2012}%
  \BibitemOpen
  \bibfield  {author} {\bibinfo {author} {\bibfnamefont {S.}~\bibnamefont
  {Bonetti}}, \bibinfo {author} {\bibfnamefont {V.}~\bibnamefont {Puliafito}},
  \bibinfo {author} {\bibfnamefont {G.}~\bibnamefont {Consolo}}, \bibinfo
  {author} {\bibfnamefont {V.~S.}\ \bibnamefont {Tiberkevich}}, \bibinfo
  {author} {\bibfnamefont {A.~N.}\ \bibnamefont {Slavin}}, \ and\ \bibinfo
  {author} {\bibfnamefont {J.}~\bibnamefont {\AA{}kerman}},\ }\href {\doibase
  10.1103/PhysRevB.85.174427} {\bibfield  {journal} {\bibinfo  {journal} {Phys.
  Rev. B}\ }\textbf {\bibinfo {volume} {85}},\ \bibinfo {pages} {174427}
  (\bibinfo {year} {2012})}\BibitemShut {NoStop}%
\bibitem [{\citenamefont {Vansteenkiste}\ \emph {et~al.}(2014)\citenamefont
  {Vansteenkiste}, \citenamefont {Leliaert}, \citenamefont {Dvornik},
  \citenamefont {Helsen}, \citenamefont {Garcia-Sanchez},\ and\ \citenamefont
  {Van~Waeyenberge}}]{VansteenKiste2014}%
  \BibitemOpen
  \bibfield  {author} {\bibinfo {author} {\bibfnamefont {A.}~\bibnamefont
  {Vansteenkiste}}, \bibinfo {author} {\bibfnamefont {J.}~\bibnamefont
  {Leliaert}}, \bibinfo {author} {\bibfnamefont {M.}~\bibnamefont {Dvornik}},
  \bibinfo {author} {\bibfnamefont {M.}~\bibnamefont {Helsen}}, \bibinfo
  {author} {\bibfnamefont {F.}~\bibnamefont {Garcia-Sanchez}}, \ and\ \bibinfo
  {author} {\bibfnamefont {B.}~\bibnamefont {Van~Waeyenberge}},\ }\href
  {\doibase 10.1063/1.4899186} {\bibfield  {journal} {\bibinfo  {journal} {AIP
  Advances}\ }\textbf {\bibinfo {volume} {4}},\ \bibinfo {pages} {107133}
  (\bibinfo {year} {2014})}\BibitemShut {NoStop}%
\bibitem [{\citenamefont {Bonetti}\ \emph {et~al.}(2015)\citenamefont
  {Bonetti}, \citenamefont {Kukreja}, \citenamefont {Chen}, \citenamefont
  {Maci\`{a}}, \citenamefont {Hern\`{a}ndez}, \citenamefont {Eklund},
  \citenamefont {Backes}, \citenamefont {Frisch}, \citenamefont {Katine},
  \citenamefont {Malm}, \citenamefont {Urazhdin}, \citenamefont {Kent},
  \citenamefont {St\"{o}hr}, \citenamefont {Ohldag},\ and\ \citenamefont
  {D\"{u}rr}}]{Bonetti2015}%
  \BibitemOpen
  \bibfield  {author} {\bibinfo {author} {\bibfnamefont {S.}~\bibnamefont
  {Bonetti}}, \bibinfo {author} {\bibfnamefont {R.}~\bibnamefont {Kukreja}},
  \bibinfo {author} {\bibfnamefont {Z.}~\bibnamefont {Chen}}, \bibinfo {author}
  {\bibfnamefont {F.}~\bibnamefont {Maci\`{a}}}, \bibinfo {author}
  {\bibfnamefont {J.~M.}\ \bibnamefont {Hern\`{a}ndez}}, \bibinfo {author}
  {\bibfnamefont {A.}~\bibnamefont {Eklund}}, \bibinfo {author} {\bibfnamefont
  {D.}~\bibnamefont {Backes}}, \bibinfo {author} {\bibfnamefont
  {J.}~\bibnamefont {Frisch}}, \bibinfo {author} {\bibfnamefont
  {J.}~\bibnamefont {Katine}}, \bibinfo {author} {\bibfnamefont
  {G.}~\bibnamefont {Malm}}, \bibinfo {author} {\bibfnamefont {S.}~\bibnamefont
  {Urazhdin}}, \bibinfo {author} {\bibfnamefont {A.~D.}\ \bibnamefont {Kent}},
  \bibinfo {author} {\bibfnamefont {J.}~\bibnamefont {St\"{o}hr}}, \bibinfo
  {author} {\bibfnamefont {H.}~\bibnamefont {Ohldag}}, \ and\ \bibinfo {author}
  {\bibfnamefont {H.}~\bibnamefont {D\"{u}rr}},\ }\href {\doibase
  10.1038/ncomms9889} {\bibfield  {journal} {\bibinfo  {journal} {Nat.
  Commun.}\ }\textbf {\bibinfo {volume} {6}},\ \bibinfo {pages} {8889}
  (\bibinfo {year} {2015})}\BibitemShut {NoStop}%
\bibitem [{\citenamefont {Leliaert}\ \emph {et~al.}(2014)\citenamefont
  {Leliaert}, \citenamefont {Van~de Wiele}, \citenamefont {Vansteenkiste},
  \citenamefont {Laurson}, \citenamefont {Durin}, \citenamefont {Dupré},\ and\
  \citenamefont {Van~Waeyenberge}}]{LeliaertJAP2014a}%
  \BibitemOpen
  \bibfield  {author} {\bibinfo {author} {\bibfnamefont {J.}~\bibnamefont
  {Leliaert}}, \bibinfo {author} {\bibfnamefont {B.}~\bibnamefont {Van~de
  Wiele}}, \bibinfo {author} {\bibfnamefont {A.}~\bibnamefont {Vansteenkiste}},
  \bibinfo {author} {\bibfnamefont {L.}~\bibnamefont {Laurson}}, \bibinfo
  {author} {\bibfnamefont {G.}~\bibnamefont {Durin}}, \bibinfo {author}
  {\bibfnamefont {L.}~\bibnamefont {Dupré}}, \ and\ \bibinfo {author}
  {\bibfnamefont {B.}~\bibnamefont {Van~Waeyenberge}},\ }\href {\doibase
  10.1063/1.4854956} {\bibfield  {journal} {\bibinfo  {journal} {Journal of
  Applied Physics}\ }\textbf {\bibinfo {volume} {115}},\ \bibinfo {pages}
  {17D102} (\bibinfo {year} {2014})}\BibitemShut {NoStop}%
\bibitem [{\citenamefont {Victora}\ \emph {et~al.}(2003)\citenamefont
  {Victora}, \citenamefont {Willoughby}, \citenamefont {MacLaren},\ and\
  \citenamefont {Xue}}]{Victora2003}%
  \BibitemOpen
  \bibfield  {author} {\bibinfo {author} {\bibfnamefont {R.~H.}\ \bibnamefont
  {Victora}}, \bibinfo {author} {\bibfnamefont {S.~D.}\ \bibnamefont
  {Willoughby}}, \bibinfo {author} {\bibfnamefont {J.~M.}\ \bibnamefont
  {MacLaren}}, \ and\ \bibinfo {author} {\bibfnamefont {J.}~\bibnamefont
  {Xue}},\ }\href {\doibase 10.1109/TMAG.2003.808998} {\bibfield  {journal}
  {\bibinfo  {journal} {IEEE Transactions on Magnetics}\ }\textbf {\bibinfo
  {volume} {39}},\ \bibinfo {pages} {710} (\bibinfo {year} {2003})}\BibitemShut
  {NoStop}%
\bibitem [{\citenamefont {Leliaert}\ \emph {et~al.}(2017)\citenamefont
  {Leliaert}, \citenamefont {Mulkers}, \citenamefont {De~Clercq}, \citenamefont
  {Coene}, \citenamefont {Dvornik},\ and\ \citenamefont
  {Van~Waeyenberge}}]{Leliaert2017}%
  \BibitemOpen
  \bibfield  {author} {\bibinfo {author} {\bibfnamefont {J.}~\bibnamefont
  {Leliaert}}, \bibinfo {author} {\bibfnamefont {J.}~\bibnamefont {Mulkers}},
  \bibinfo {author} {\bibfnamefont {J.}~\bibnamefont {De~Clercq}}, \bibinfo
  {author} {\bibfnamefont {A.}~\bibnamefont {Coene}}, \bibinfo {author}
  {\bibfnamefont {M.}~\bibnamefont {Dvornik}}, \ and\ \bibinfo {author}
  {\bibfnamefont {B.}~\bibnamefont {Van~Waeyenberge}},\ }\href {\doibase
  10.1063/1.5003957} {\bibfield  {journal} {\bibinfo  {journal} {AIP Advances}\
  }\textbf {\bibinfo {volume} {7}},\ \bibinfo {pages} {125010} (\bibinfo {year}
  {2017})}\BibitemShut {NoStop}%
\bibitem [{\citenamefont {Iacocca}\ \emph {et~al.}(2015)\citenamefont
  {Iacocca}, \citenamefont {D\"urrenfeld}, \citenamefont {Heinonen},
  \citenamefont {\AA{}kerman},\ and\ \citenamefont {Dumas}}]{Iacocca2015}%
  \BibitemOpen
  \bibfield  {author} {\bibinfo {author} {\bibfnamefont {E.}~\bibnamefont
  {Iacocca}}, \bibinfo {author} {\bibfnamefont {P.}~\bibnamefont
  {D\"urrenfeld}}, \bibinfo {author} {\bibfnamefont {O.}~\bibnamefont
  {Heinonen}}, \bibinfo {author} {\bibfnamefont {J.}~\bibnamefont
  {\AA{}kerman}}, \ and\ \bibinfo {author} {\bibfnamefont {R.~K.}\ \bibnamefont
  {Dumas}},\ }\href {\doibase 10.1103/PhysRevB.91.104405} {\bibfield  {journal}
  {\bibinfo  {journal} {Phys. Rev. B}\ }\textbf {\bibinfo {volume} {91}},\
  \bibinfo {pages} {104405} (\bibinfo {year} {2015})}\BibitemShut {NoStop}%
\bibitem [{\citenamefont {{Puliafito}}\ \emph {et~al.}(2014)\citenamefont
  {{Puliafito}}, \citenamefont {{Pogoryelov}}, \citenamefont {{Azzerboni}},
  \citenamefont {{{\AA}kerman}},\ and\ \citenamefont
  {{Finocchio}}}]{Puliafito2014}%
  \BibitemOpen
  \bibfield  {author} {\bibinfo {author} {\bibfnamefont {V.}~\bibnamefont
  {{Puliafito}}}, \bibinfo {author} {\bibfnamefont {Y.}~\bibnamefont
  {{Pogoryelov}}}, \bibinfo {author} {\bibfnamefont {B.}~\bibnamefont
  {{Azzerboni}}}, \bibinfo {author} {\bibfnamefont {J.}~\bibnamefont
  {{{\AA}kerman}}}, \ and\ \bibinfo {author} {\bibfnamefont {G.}~\bibnamefont
  {{Finocchio}}},\ }\href {\doibase 10.1109/TNANO.2014.2308474} {\bibfield
  {journal} {\bibinfo  {journal} {IEEE Transactions on Nanotechnology}\
  }\textbf {\bibinfo {volume} {13}},\ \bibinfo {pages} {532} (\bibinfo {year}
  {2014})}\BibitemShut {NoStop}%
\bibitem [{\citenamefont {Rippard}\ \emph {et~al.}(2005)\citenamefont
  {Rippard}, \citenamefont {Pufall}, \citenamefont {Kaka}, \citenamefont
  {Silva}, \citenamefont {Russek},\ and\ \citenamefont {Katine}}]{Rippard2005}%
  \BibitemOpen
  \bibfield  {author} {\bibinfo {author} {\bibfnamefont {W.~H.}\ \bibnamefont
  {Rippard}}, \bibinfo {author} {\bibfnamefont {M.~R.}\ \bibnamefont {Pufall}},
  \bibinfo {author} {\bibfnamefont {S.}~\bibnamefont {Kaka}}, \bibinfo {author}
  {\bibfnamefont {T.~J.}\ \bibnamefont {Silva}}, \bibinfo {author}
  {\bibfnamefont {S.~E.}\ \bibnamefont {Russek}}, \ and\ \bibinfo {author}
  {\bibfnamefont {J.~A.}\ \bibnamefont {Katine}},\ }\href {\doibase
  10.1103/PhysRevLett.95.067203} {\bibfield  {journal} {\bibinfo  {journal}
  {Phys. Rev. Lett.}\ }\textbf {\bibinfo {volume} {95}},\ \bibinfo {pages}
  {067203} (\bibinfo {year} {2005})}\BibitemShut {NoStop}%
\bibitem [{\citenamefont {Florez}\ \emph {et~al.}(2008)\citenamefont {Florez},
  \citenamefont {Katine}, \citenamefont {Carey}, \citenamefont {Folks},
  \citenamefont {Ozatay},\ and\ \citenamefont {Terris}}]{Florez2008}%
  \BibitemOpen
  \bibfield  {author} {\bibinfo {author} {\bibfnamefont {S.~H.}\ \bibnamefont
  {Florez}}, \bibinfo {author} {\bibfnamefont {J.~A.}\ \bibnamefont {Katine}},
  \bibinfo {author} {\bibfnamefont {M.}~\bibnamefont {Carey}}, \bibinfo
  {author} {\bibfnamefont {L.}~\bibnamefont {Folks}}, \bibinfo {author}
  {\bibfnamefont {O.}~\bibnamefont {Ozatay}}, \ and\ \bibinfo {author}
  {\bibfnamefont {B.~D.}\ \bibnamefont {Terris}},\ }\href {\doibase
  10.1103/PhysRevB.78.184403} {\bibfield  {journal} {\bibinfo  {journal} {Phys.
  Rev. B}\ }\textbf {\bibinfo {volume} {78}},\ \bibinfo {pages} {184403}
  (\bibinfo {year} {2008})}\BibitemShut {NoStop}%
\bibitem [{\citenamefont {Urazhdin}\ \emph {et~al.}(2010)\citenamefont
  {Urazhdin}, \citenamefont {Tabor}, \citenamefont {Tiberkevich},\ and\
  \citenamefont {Slavin}}]{Urazhdin2010}%
  \BibitemOpen
  \bibfield  {author} {\bibinfo {author} {\bibfnamefont {S.}~\bibnamefont
  {Urazhdin}}, \bibinfo {author} {\bibfnamefont {P.}~\bibnamefont {Tabor}},
  \bibinfo {author} {\bibfnamefont {V.}~\bibnamefont {Tiberkevich}}, \ and\
  \bibinfo {author} {\bibfnamefont {A.}~\bibnamefont {Slavin}},\ }\href
  {\doibase 10.1103/PhysRevLett.105.104101} {\bibfield  {journal} {\bibinfo
  {journal} {Phys. Rev. Lett.}\ }\textbf {\bibinfo {volume} {105}},\ \bibinfo
  {pages} {104101} (\bibinfo {year} {2010})}\BibitemShut {NoStop}%
\bibitem [{\citenamefont {Quinsat}\ \emph {et~al.}(2011)\citenamefont
  {Quinsat}, \citenamefont {Sierra}, \citenamefont {Firastrau}, \citenamefont
  {Tiberkevich}, \citenamefont {Slavin}, \citenamefont {Gusakova},
  \citenamefont {Buda-Prejbeanu}, \citenamefont {Zarudniev}, \citenamefont
  {Michel}, \citenamefont {Ebels}, \citenamefont {Dieny}, \citenamefont
  {Cyrille}, \citenamefont {Katine}, \citenamefont {Mauri},\ and\ \citenamefont
  {Zeltser}}]{Quinsat2011}%
  \BibitemOpen
  \bibfield  {author} {\bibinfo {author} {\bibfnamefont {M.}~\bibnamefont
  {Quinsat}}, \bibinfo {author} {\bibfnamefont {J.~F.}\ \bibnamefont {Sierra}},
  \bibinfo {author} {\bibfnamefont {I.}~\bibnamefont {Firastrau}}, \bibinfo
  {author} {\bibfnamefont {V.}~\bibnamefont {Tiberkevich}}, \bibinfo {author}
  {\bibfnamefont {A.}~\bibnamefont {Slavin}}, \bibinfo {author} {\bibfnamefont
  {D.}~\bibnamefont {Gusakova}}, \bibinfo {author} {\bibfnamefont {L.~D.}\
  \bibnamefont {Buda-Prejbeanu}}, \bibinfo {author} {\bibfnamefont
  {M.}~\bibnamefont {Zarudniev}}, \bibinfo {author} {\bibfnamefont {J.-P.}\
  \bibnamefont {Michel}}, \bibinfo {author} {\bibfnamefont {U.}~\bibnamefont
  {Ebels}}, \bibinfo {author} {\bibfnamefont {B.}~\bibnamefont {Dieny}},
  \bibinfo {author} {\bibfnamefont {M.-C.}\ \bibnamefont {Cyrille}}, \bibinfo
  {author} {\bibfnamefont {J.~A.}\ \bibnamefont {Katine}}, \bibinfo {author}
  {\bibfnamefont {D.}~\bibnamefont {Mauri}}, \ and\ \bibinfo {author}
  {\bibfnamefont {A.}~\bibnamefont {Zeltser}},\ }\href {\doibase
  http://dx.doi.org/10.1063/1.3587575} {\bibfield  {journal} {\bibinfo
  {journal} {Appl. Phys. Lett.}\ }\textbf {\bibinfo {volume} {98}},\ \bibinfo
  {eid} {182503} (\bibinfo {year} {2011})}\BibitemShut {NoStop}%
\bibitem [{\citenamefont {{Muduli}}\ \emph {et~al.}(2011)\citenamefont
  {{Muduli}}, \citenamefont {{Pogoryelov}}, \citenamefont {{Zhou}},
  \citenamefont {{Mancoff}},\ and\ \citenamefont
  {{{\AA}kerman}}}]{muduli2011if}%
  \BibitemOpen
  \bibfield  {author} {\bibinfo {author} {\bibfnamefont {P.~K.}\ \bibnamefont
  {{Muduli}}}, \bibinfo {author} {\bibfnamefont {Y.}~\bibnamefont
  {{Pogoryelov}}}, \bibinfo {author} {\bibfnamefont {Y.}~\bibnamefont
  {{Zhou}}}, \bibinfo {author} {\bibfnamefont {F.}~\bibnamefont {{Mancoff}}}, \
  and\ \bibinfo {author} {\bibfnamefont {J.}~\bibnamefont {{{\AA}kerman}}},\
  }\href {\doibase 10.1080/10584587.2011.574478} {\bibfield  {journal}
  {\bibinfo  {journal} {Integr. Ferroelectr.}\ }\textbf {\bibinfo {volume}
  {125}},\ \bibinfo {pages} {147} (\bibinfo {year} {2011})}\BibitemShut
  {NoStop}%
\bibitem [{\citenamefont {Finocchio}\ \emph {et~al.}(2012)\citenamefont
  {Finocchio}, \citenamefont {Carpentieri}, \citenamefont {Giordano},\ and\
  \citenamefont {Azzerboni}}]{Finocchio2012}%
  \BibitemOpen
  \bibfield  {author} {\bibinfo {author} {\bibfnamefont {G.}~\bibnamefont
  {Finocchio}}, \bibinfo {author} {\bibfnamefont {M.}~\bibnamefont
  {Carpentieri}}, \bibinfo {author} {\bibfnamefont {A.}~\bibnamefont
  {Giordano}}, \ and\ \bibinfo {author} {\bibfnamefont {B.}~\bibnamefont
  {Azzerboni}},\ }\href {\doibase 10.1103/PhysRevB.86.014438} {\bibfield
  {journal} {\bibinfo  {journal} {Phys. Rev. B}\ }\textbf {\bibinfo {volume}
  {86}},\ \bibinfo {pages} {014438} (\bibinfo {year} {2012})}\BibitemShut
  {NoStop}%
\bibitem [{\citenamefont {Kaka}\ \emph {et~al.}(2005)\citenamefont {Kaka},
  \citenamefont {Pufall}, \citenamefont {Rippard}, \citenamefont {Silva},
  \citenamefont {Russek},\ and\ \citenamefont {Katine}}]{Kaka2005}%
  \BibitemOpen
  \bibfield  {author} {\bibinfo {author} {\bibfnamefont {S.}~\bibnamefont
  {Kaka}}, \bibinfo {author} {\bibfnamefont {M.~R.}\ \bibnamefont {Pufall}},
  \bibinfo {author} {\bibfnamefont {W.~H.}\ \bibnamefont {Rippard}}, \bibinfo
  {author} {\bibfnamefont {T.~J.}\ \bibnamefont {Silva}}, \bibinfo {author}
  {\bibfnamefont {S.~E.}\ \bibnamefont {Russek}}, \ and\ \bibinfo {author}
  {\bibfnamefont {J.~A.}\ \bibnamefont {Katine}},\ }\href {\doibase
  10.1038/nature04035} {\bibfield  {journal} {\bibinfo  {journal} {Nature}\
  }\textbf {\bibinfo {volume} {437}},\ \bibinfo {pages} {389} (\bibinfo {year}
  {2005})}\BibitemShut {NoStop}%
\bibitem [{\citenamefont {Mancoff}\ \emph {et~al.}(2005)\citenamefont
  {Mancoff}, \citenamefont {Rizzo}, \citenamefont {Engel},\ and\ \citenamefont
  {Tehrani}}]{Mancoff2005}%
  \BibitemOpen
  \bibfield  {author} {\bibinfo {author} {\bibfnamefont {F.~B.}\ \bibnamefont
  {Mancoff}}, \bibinfo {author} {\bibfnamefont {N.~D.}\ \bibnamefont {Rizzo}},
  \bibinfo {author} {\bibfnamefont {B.~N.}\ \bibnamefont {Engel}}, \ and\
  \bibinfo {author} {\bibfnamefont {S.}~\bibnamefont {Tehrani}},\ }\href
  {\doibase 10.1038/nature04036} {\bibfield  {journal} {\bibinfo  {journal}
  {Nature}\ }\textbf {\bibinfo {volume} {437}},\ \bibinfo {pages} {393}
  (\bibinfo {year} {2005})}\BibitemShut {NoStop}%
\bibitem [{\citenamefont {Pufall}\ \emph {et~al.}(2006)\citenamefont {Pufall},
  \citenamefont {Rippard}, \citenamefont {Russek}, \citenamefont {Kaka},\ and\
  \citenamefont {Katine}}]{Pufall2006}%
  \BibitemOpen
  \bibfield  {author} {\bibinfo {author} {\bibfnamefont {M.~R.}\ \bibnamefont
  {Pufall}}, \bibinfo {author} {\bibfnamefont {W.~H.}\ \bibnamefont {Rippard}},
  \bibinfo {author} {\bibfnamefont {S.~E.}\ \bibnamefont {Russek}}, \bibinfo
  {author} {\bibfnamefont {S.}~\bibnamefont {Kaka}}, \ and\ \bibinfo {author}
  {\bibfnamefont {J.~A.}\ \bibnamefont {Katine}},\ }\href {\doibase
  10.1103/PhysRevLett.97.087206} {\bibfield  {journal} {\bibinfo  {journal}
  {Phys. Rev. Lett.}\ }\textbf {\bibinfo {volume} {97}},\ \bibinfo {pages}
  {087206} (\bibinfo {year} {2006})}\BibitemShut {NoStop}%
\bibitem [{\citenamefont {Rezende}\ \emph {et~al.}(2007)\citenamefont
  {Rezende}, \citenamefont {de~Aguiar}, \citenamefont
  {Rodr\'{\i}guez-Su\'arez},\ and\ \citenamefont {Azevedo}}]{Rezende2007}%
  \BibitemOpen
  \bibfield  {author} {\bibinfo {author} {\bibfnamefont {S.~M.}\ \bibnamefont
  {Rezende}}, \bibinfo {author} {\bibfnamefont {F.~M.}\ \bibnamefont
  {de~Aguiar}}, \bibinfo {author} {\bibfnamefont {R.~L.}\ \bibnamefont
  {Rodr\'{\i}guez-Su\'arez}}, \ and\ \bibinfo {author} {\bibfnamefont
  {A.}~\bibnamefont {Azevedo}},\ }\href {\doibase
  10.1103/PhysRevLett.98.087202} {\bibfield  {journal} {\bibinfo  {journal}
  {Phys. Rev. Lett.}\ }\textbf {\bibinfo {volume} {98}},\ \bibinfo {pages}
  {087202} (\bibinfo {year} {2007})}\BibitemShut {NoStop}%
\bibitem [{\citenamefont {Sani}\ \emph {et~al.}(2013)\citenamefont {Sani},
  \citenamefont {Persson}, \citenamefont {Mohseni}, \citenamefont {Pogoryelov},
  \citenamefont {Muduli}, \citenamefont {Eklund}, \citenamefont {Malm},
  \citenamefont {K\"all}, \citenamefont {Dmitriev},\ and\ \citenamefont
  {{\AA}kerman}}]{Sani2013a}%
  \BibitemOpen
  \bibfield  {author} {\bibinfo {author} {\bibfnamefont {S.}~\bibnamefont
  {Sani}}, \bibinfo {author} {\bibfnamefont {J.}~\bibnamefont {Persson}},
  \bibinfo {author} {\bibfnamefont {S.}~\bibnamefont {Mohseni}}, \bibinfo
  {author} {\bibfnamefont {Y.}~\bibnamefont {Pogoryelov}}, \bibinfo {author}
  {\bibfnamefont {P.}~\bibnamefont {Muduli}}, \bibinfo {author} {\bibfnamefont
  {A.}~\bibnamefont {Eklund}}, \bibinfo {author} {\bibfnamefont
  {G.}~\bibnamefont {Malm}}, \bibinfo {author} {\bibfnamefont {M.}~\bibnamefont
  {K\"all}}, \bibinfo {author} {\bibfnamefont {A.}~\bibnamefont {Dmitriev}}, \
  and\ \bibinfo {author} {\bibfnamefont {J.}~\bibnamefont {{\AA}kerman}},\
  }\href {\doibase 10.1038/ncomms3731} {\bibfield  {journal} {\bibinfo
  {journal} {Nat. Commun.}\ }\textbf {\bibinfo {volume} {4}},\ \bibinfo {pages}
  {2731} (\bibinfo {year} {2013})}\BibitemShut {NoStop}%
\bibitem [{\citenamefont {Houshang}\ \emph {et~al.}(2016)\citenamefont
  {Houshang}, \citenamefont {Iacocca}, \citenamefont {D{\"u}rrenfeld},
  \citenamefont {Sani}, \citenamefont {{\AA}kerman},\ and\ \citenamefont
  {Dumas}}]{Houshang2015}%
  \BibitemOpen
  \bibfield  {author} {\bibinfo {author} {\bibfnamefont {A.}~\bibnamefont
  {Houshang}}, \bibinfo {author} {\bibfnamefont {E.}~\bibnamefont {Iacocca}},
  \bibinfo {author} {\bibfnamefont {P.}~\bibnamefont {D{\"u}rrenfeld}},
  \bibinfo {author} {\bibfnamefont {S.~R.}\ \bibnamefont {Sani}}, \bibinfo
  {author} {\bibfnamefont {J.}~\bibnamefont {{\AA}kerman}}, \ and\ \bibinfo
  {author} {\bibfnamefont {R.~K.}\ \bibnamefont {Dumas}},\ }\href {\doibase
  10.1038/nnano.2015.280} {\bibfield  {journal} {\bibinfo  {journal} {Nature
  Nanotechnology}\ }\textbf {\bibinfo {volume} {11}},\ \bibinfo {pages} {280}
  (\bibinfo {year} {2016})}\BibitemShut {NoStop}%
\bibitem [{\citenamefont {Banuazizi}\ \emph {et~al.}(2017)\citenamefont
  {Banuazizi}, \citenamefont {Sani}, \citenamefont {Eklund}, \citenamefont
  {Naiini}, \citenamefont {Mohseni}, \citenamefont {Chung}, \citenamefont
  {D\"urrenfeld}, \citenamefont {Malm},\ and\ \citenamefont
  {{\AA}kerman}}]{Banuazizi2017}%
  \BibitemOpen
  \bibfield  {author} {\bibinfo {author} {\bibfnamefont {S.~A.~H.}\
  \bibnamefont {Banuazizi}}, \bibinfo {author} {\bibfnamefont {S.~R.}\
  \bibnamefont {Sani}}, \bibinfo {author} {\bibfnamefont {A.}~\bibnamefont
  {Eklund}}, \bibinfo {author} {\bibfnamefont {M.~M.}\ \bibnamefont {Naiini}},
  \bibinfo {author} {\bibfnamefont {S.~M.}\ \bibnamefont {Mohseni}}, \bibinfo
  {author} {\bibfnamefont {S.}~\bibnamefont {Chung}}, \bibinfo {author}
  {\bibfnamefont {P.}~\bibnamefont {D\"urrenfeld}}, \bibinfo {author}
  {\bibfnamefont {B.~G.}\ \bibnamefont {Malm}}, \ and\ \bibinfo {author}
  {\bibfnamefont {J.}~\bibnamefont {{\AA}kerman}},\ }\href {\doibase
  10.1039/C6NR07309C} {\bibfield  {journal} {\bibinfo  {journal} {Nanoscale}\
  }\textbf {\bibinfo {volume} {9}},\ \bibinfo {pages} {1896} (\bibinfo {year}
  {2017})}\BibitemShut {NoStop}%
\bibitem [{\citenamefont {Muduli}\ \emph {et~al.}(2012)\citenamefont {Muduli},
  \citenamefont {Heinonen},\ and\ \citenamefont {\AA{}kerman}}]{Muduli2012b}%
  \BibitemOpen
  \bibfield  {author} {\bibinfo {author} {\bibfnamefont {P.~K.}\ \bibnamefont
  {Muduli}}, \bibinfo {author} {\bibfnamefont {O.~G.}\ \bibnamefont
  {Heinonen}}, \ and\ \bibinfo {author} {\bibfnamefont {J.}~\bibnamefont
  {\AA{}kerman}},\ }\href {\doibase 10.1103/PhysRevB.86.174408} {\bibfield
  {journal} {\bibinfo  {journal} {Phys. Rev. B}\ }\textbf {\bibinfo {volume}
  {86}},\ \bibinfo {pages} {174408} (\bibinfo {year} {2012})}\BibitemShut
  {NoStop}%
\bibitem [{\citenamefont {Eklund}\ \emph {et~al.}(2013)\citenamefont {Eklund},
  \citenamefont {Sani}, \citenamefont {Mohseni}, \citenamefont {Persson},
  \citenamefont {Malm},\ and\ \citenamefont {{\AA}kerman}}]{Eklund2013}%
  \BibitemOpen
  \bibfield  {author} {\bibinfo {author} {\bibfnamefont {A.~J.}\ \bibnamefont
  {Eklund}}, \bibinfo {author} {\bibfnamefont {S.~R.}\ \bibnamefont {Sani}},
  \bibinfo {author} {\bibfnamefont {S.~M.}\ \bibnamefont {Mohseni}}, \bibinfo
  {author} {\bibfnamefont {J.}~\bibnamefont {Persson}}, \bibinfo {author}
  {\bibfnamefont {B.~G.}\ \bibnamefont {Malm}}, \ and\ \bibinfo {author}
  {\bibfnamefont {J.}~\bibnamefont {{\AA}kerman}},\ }in\ \href {\doibase
  10.1109/ICNF.2013.6578965} {\emph {\bibinfo {booktitle} {2013 22nd
  International Conference on Noise and Fluctuations (ICNF)}}}\ (\bibinfo
  {year} {2013})\ pp.\ \bibinfo {pages} {1--4},\ \bibinfo {note} {{DOI:
  10.1109/ICNF.2013.6578965}}\BibitemShut {NoStop}%
\bibitem [{\citenamefont {Malm}\ \emph {et~al.}(2019)\citenamefont {Malm},
  \citenamefont {Eklund},\ and\ \citenamefont {Dvornik}}]{Malm2019}%
  \BibitemOpen
  \bibfield  {author} {\bibinfo {author} {\bibfnamefont {B.~G.}\ \bibnamefont
  {Malm}}, \bibinfo {author} {\bibfnamefont {A.}~\bibnamefont {Eklund}}, \ and\
  \bibinfo {author} {\bibfnamefont {M.}~\bibnamefont {Dvornik}},\ }\href@noop
  {} {\enquote {\bibinfo {title} {Micromagnetic modeling of telegraphic mode
  jumping in microwave spin torque oscillators},}\ } (\bibinfo {year} {2019}),\
  \Eprint {http://arxiv.org/abs/1909.08431} {arXiv:1909.08431
  [cond-mat.mes-hall]} \BibitemShut {NoStop}%
\bibitem [{\citenamefont {Heinonen}\ \emph {et~al.}(2013)\citenamefont
  {Heinonen}, \citenamefont {Muduli}, \citenamefont {Iacocca},\ and\
  \citenamefont {{\AA}kerman}}]{Heinonen2013}%
  \BibitemOpen
  \bibfield  {author} {\bibinfo {author} {\bibfnamefont {O.}~\bibnamefont
  {Heinonen}}, \bibinfo {author} {\bibfnamefont {P.}~\bibnamefont {Muduli}},
  \bibinfo {author} {\bibfnamefont {E.}~\bibnamefont {Iacocca}}, \ and\
  \bibinfo {author} {\bibfnamefont {J.}~\bibnamefont {{\AA}kerman}},\ }\href
  {\doibase 10.1109/TMAG.2013.2242866} {\bibfield  {journal} {\bibinfo
  {journal} {IEEE Trans. Magn.}\ }\textbf {\bibinfo {volume} {49}},\ \bibinfo
  {pages} {4398} (\bibinfo {year} {2013})}\BibitemShut {NoStop}%
\end{thebibliography}%

\end{document}